\documentclass{aa}
\usepackage{epsf}
\usepackage{times}
\usepackage{graphics}
\newcommand{\etal}{et al.} 

\begin{document}
\thesaurus{12         
          (02.05.2;   
           02.16.1;   
           08.09.3;   
           08.05.3;   
           08.12.2;   
           08.23.1)   
          }
\title{Thermodynamical properties of stellar matter}

\subtitle{II. Internal energy, temperature and density exponents, \\ and
              specific heats for stellar interiors}
              
\author{W.\, Stolzmann\inst{1,2} and T.\, Bl\"ocker\inst{3}}   
\offprints{W.\, Stolzmann}
\institute
{Institut f\"ur Theoretische Physik und Astrophysik, Olshausenstr.\ 40,
  D-24098 Kiel, Germany
  (stolzmann@astrophysik.uni-kiel.de)
\and
  Astrophysikalisches Institut Potsdam, An der Sternwarte 16,
  D-14482 Potsdam, Germany
\and 
  Max-Planck-Institut f\"ur Radioastronomie, Auf dem H\"ugel 69,
  D-53121 Bonn, Germany (bloecker@mpifr-bonn.mpg.de)}

\date{Received date; accepted date}
 
\maketitle
 
\begin{abstract}
Starting from the Helmholtz free energy we calculate analytically 
first- and second-order derivatives, as internal energy and specific heats, 
for the ideal system and the exchange and correlation interactions covering 
a broad range of degeneracy and relativity.
The complex physics of Coulomb interactions is expressed by Pad\'e 
Approximants, which reflect the actual state of our knowledge with high 
accuracy. We assume complete ionization and provide a base system of 
thermodynamical functions from which any other thermodynamical quantities 
can be calculated.
We chose for the base system  the free energy, the pressure, the internal 
energy, the isothermal compressibility (or density exponent), 
the coefficient of strain (or temperature exponent), 
and the isochoric specific heat. By means of the latter 
potentials entropy, isobaric specific heat and adiabatic temperature gradient 
can be determined. 
We give comparisons with quantities which are composed by numerical
second-order derivatives of the free energy and show that numerical 
derivatives of the free energy as calculated, for instance, from EOS tables, 
may produce discontinuities for astrophysically relevant quantities as,
e.g., the adiabatic temperature gradient.
Adiabatic temperature gradients are shown for different chemical 
compositions (hydrogen, helium, carbon). 
Finally the used formalism of  Pad\'e Approximants allows immediate 
incorporation of recent results from many particle statistics. 
\end{abstract}
\section{Introduction} \label{intro}
For an accurate modelling of stellar objects we have to compute a complete set 
of thermodynamical quantities which meets the physical conditions of various 
evolutionary stages.
Based on the framework presented in Stolzmann \& Bl\"ocker (1996a, hereafter
paper I) we derive further thermodynamical potentials in order to provide a
base system from which {\it any} other thermodynamical quantity can
be calculated. Such a system requires six quantities consisting of
the Helmholtz free energy, two of its single derivatives, one of its second 
mixed derivatives, and two of its second pure derivatives with respect to   
temperature and density (see e.g. D\"appen \etal\ 1988). 
We complement the Helmholtz free energy with the
pressure and the internal energy, with the isochoric specific heat, and
with the temperature and density exponents in the equation of state (EOS),
$\chi_{\rm T}$ and $\chi_{\rho}$.
This paper will be devoted to the ideal, exchange and correlation 
contributions according to the 
expansions given in paper I, viz.\ for
{\it case a}: weak relativity and arbitrary degeneracy, and {\it case b}:
strong degeneracy and arbitrary relativity. These cases cover a large
area in the density-temperature plane. 
The correlations between the charged particles are formulated
by the technique of Pad\'e Approximants as in paper I.
Here we give improved versions of our earlier applied Pad\'e Approximants.
In particular we have rearranged terms of the quantum virial function.
\\
Explicit expressions for the Helmholtz free energy and the
pressure have already been derived in paper I for ideality as well as for
exchange and correlation interaction.
This paper aims at providing analytical formulae with high accuracy 
which supply quick computing with reliable accuracy in practical applications.
Furthermore the expressions of the thermodynamical functions given here 
can be easily included as a part of the thermodynamical description of 
arbitrary degree of ionization.
\\
This paper is organized as follows. In Sect.~\ref{thermrel} 
we list the set of 
thermodynamical potentials to be considered. Sect.~\ref{theomodel} gives a 
brief overview of the concept for the calculation of the EOS terms. 
Sect.~\ref{thermppot} deals with the detailed determination of the ideal, 
exchange and correlation parts for the EOS. 
Numerical results and comparisons are presented in Sect.~\ref{results}, 
and a summary is given in Sect.~\ref{summ}.
\section{Thermodynamical relations and identities}    \label{thermrel}
We summarize briefly some well-known standard relations, which are frequently 
used to provide the thermodynamics for astrophysical applications.
\\
We have to determine the first and second-order quantities which are
related by the ratio $C_{\rm P}$/$C_{\rm V}$ of isochoric to isobaric 
specific heat:
\begin{equation}  \label{Egamspez}
\gamma = \frac{C_{\rm P}}{C_{\rm V}} = 1 + \frac{ V }{ K_{\rm T}} \;
                                           \frac{\lambda_{\rm P}^{2}}   
                                                {C_{\rm V} T } \; ,
\end{equation}
The (isobaric) thermal expansion coefficient
$\lambda_{\rm P}$ can be expressed by
\begin{equation}
\lambda_{\rm P} = \frac{K_{\rm T}}{V} C_{\rm V} T \gamma_{\rm G}
                = K_{\rm T}\Phi_{\rm S}
 \;\;   ,
\end{equation}
with $\gamma_{\rm G}$ being the Gr\"uneisen coefficient
\begin{equation}
\gamma_{\rm G} = \frac{P V}{C_{\rm V} T } \chi_{\rm T} \;\;  .
\end{equation}
$\Phi_{\rm S}$ is the coefficient of strain 
\begin{equation}   \label{EPhi}
\Phi_{\rm S} = P \chi_{\rm T}  \;\;  ,
\end{equation}
and $K_{\rm T}$ is the isothermal compressibility
\begin{equation}   \label{Ecomp}
\frac {1}{K_{\rm T}} = P \chi_{\rho} \;\;  .
\end{equation}               
The so-called temperature and density exponents in the equation of
state (Cox \& Giuli 1968) are defined by
\begin{equation}  \label{EchiT}
\chi_{\rm T} = \left( \frac{\partial \ln P}{\partial \ln T} \right)_{\rho}
             = \frac{T}{P} \; \left( \frac{\partial P}{\partial T} \right)_{V}
\end{equation}
\begin{equation}  \label{EchiR}
\chi_{\rho} = \left( \frac{\partial \ln P}{\partial \ln \rho} \right)_{T}
            = - \frac{V}{P} \; \left( \frac{\partial P}{\partial V} \right)_{T}
   \;\;  .
\end{equation}
The adiabatic temperature gradient defined by
$\nabla_{\rm ad}=(\partial \ln T/\partial \ln P)_{S}$
($S$ denotes the entropy)
can be expressed by
\begin{equation}  \label{Enabla}
\nabla_{\rm ad} = \frac{P V}{C_{\rm P} T} \lambda_{\rm P}
                = \frac{P V}{C_{\rm P}T}\frac{\chi_{\rm T}}{\chi_{\rho}} \;\; . 
\end{equation}  
Another possibility to calculate the adiabatic gradient is given by 
the three adiabatic exponents $\Gamma_{1}$, $\Gamma_{2}$, 
and $\Gamma_{3}$ (Cox \& Giuli 1968, Rogers \etal\ 1996)
\begin{equation}
\nabla_{\rm ad} = \frac{\Gamma_{2}-1}{\Gamma_{2}}     
                = \frac{\Gamma_{3}-1}{\Gamma_{1}}  \;\; ,
\end{equation}
which can be obtained by
\begin{equation} \label{EGams}
\Gamma_{1} = \frac{\gamma}{P K_{\rm T}} \; , \quad \;\;
\Gamma_{2} = \frac{\gamma}{\gamma-P K_{\rm T}\gamma_{\rm G}}  \; , \quad \;\;
\Gamma_{3} = 1 + \gamma_{\rm G} \; .
\end{equation}
In order to calculate the quantities given 
by Eqs.~(\ref{Egamspez})-(\ref{EGams})
we start with the Helmholtz free energy $F(V,T,N)$ and determine 
by means of standard thermodynamic relations the Gibbs energy,
\begin{equation}
G = N \left( \frac{\partial F}{\partial N} \right)_{T,V}  \;\; ,
\end{equation}
the pressure,
\begin{equation}  \label{EPbas}
PV = G - F  \;\, ,
\end{equation}
and the internal energy,
\begin{equation} \label{EUbas}
U = F - T \left( \frac{\partial F}{\partial T} \right)_{V,N}          \;\;,
\end{equation}
leading to the  entropy $S$ according to
\begin{equation}
T S  = U - F
\end{equation}
and to the isochoric specific heat via
\begin{equation}  \label{ECbas}
C_{\rm V} = \left( \frac{\partial U}{\partial T} \right)_{V,N}
  = - T  \left( \frac{\partial^{2} F}{\partial T^{2}} \right)_{V,N}  \;\; .
\end{equation}
Moreover, we have to calculate the coefficient of strain and the inverse 
compressibility (bulk modulus) by means of Eqs.~(\ref{EPhi})-(\ref{EchiR}).
\section{Theoretical model}  \label{theomodel}
As in paper I we start with the Helmholtz free energy $F$ of a fully
ionized plasma consisting of ideal and Coulomb interaction parts
\begin{equation}  \label{Eics}
F(T,V,N_{a}) = \sum_{a} F_{a}^{\rm id} + F^{\rm coul} \;\;\;\; ,  
\end{equation}
where the Coulomb term contains the following parts:
\begin{equation}  \label{Ecos}
F^{\rm coul} = F_{\rm ee}^{\rm x} + F_{\rm ee}^{\rm c} + F_{\rm ii}^{\rm c} +
               F_{\rm ii}^{\rm cq} + F_{\rm ie}^{\rm c} \;\;\;\; ,
\end{equation}
with x and c marking the exchange and correlation contributions and cq 
the quantum correction term.
The pairs $ab$ denote the interaction between particles of species
$a$ and $b$ (electrons, ions), respectively.

For convenience we introduce dimensionless thermodynamical potentials 
defined by 
\begin{equation} \label{Efdim1} 
f = \frac{F}{NkT} \; , \quad \;\;  
g = \frac{G}{NkT} \; , \quad \;\; 
p = \frac{P}{nkT} \; , \quad \;\;  
u = \frac{U}{NkT}  
\end{equation}
\begin{equation} \label{Efdim2}  
s = \frac{S}{Nk} \; , \quad 
\frac{1}{k_{\rm T}} = \frac{1}{K_{\rm T} \; nkT} \; , \quad 
\phi_{\rm S} =        \frac{\Phi_{\rm S}}  {nkT} \; , \quad 
c_{\rm V}= \frac{C_{\rm V}}{Nk} 
\end{equation}
where $n = N/V$ refers to the total particle number density of the ions 
{\it or} the electrons, and $k$ denotes the Boltzmann constant.
Eq.~(\ref{Eics}), or correspondingly the potentials summarized by
Eq.~(\ref{Efdim1}) and Eq.~(\ref{Efdim2}), can be written by    
\begin{eqnarray}  \label{Eicsredu}
\Sigma & = & N_{\rm e}kT \left[\sigma_{\rm e }^{\rm id}
                             + \sigma_{\rm ee}^{\rm x } 
                             + \sigma_{\rm ee}^{\rm c } \right]  + 
 \nonumber \\ 
 & &         N_{\rm i}kT \left[\sigma_{\rm i }^{\rm id} 
                             + \sigma_{\rm ii}^{\rm c }
                             + \sigma_{\rm ii}^{\rm cq}    
                             + \sigma_{\rm ie}^{\rm c } \right]
\end{eqnarray}
where 
$\Sigma = \{ F,\; G, \; P \cdot V, \; U, \; S \cdot T, \; V/K_{\rm T}, \; 
\Phi_{\rm S} \cdot V, \; C_{\rm V} \cdot T \}$ and
$\sigma=\{ f, g, p, u, s, 1/k_{\rm T}, \phi_{\rm S}, c_{\rm V} \}$
symbolize the various thermodynamic functions defined in the previous section.
\\
\section{Thermodynamical potentials}   \label{thermppot}
This section deals in detail with the calculation of the potentials listed in 
Eq.~(\ref{Eicsredu}).
By introducing definitions and for the sake of the integrity we repeat here 
few expressions concerning Helmholtz free energy and pressure, which are
already given in paper I.  
\subsection{Ideality}
The ideality of the nonrelativistic and nondegenerate ions is described 
by the well-known classical expressions
for the Helmholtz free energy 
\begin{equation}  \label{Eionid}
f_{\rm i}^{\rm id} = 
   \left[ \ln \left( n_{\rm i} \Lambda_{\rm i} ^{3} \right) - 1 \right] \;\; .
\end{equation}
The thermal de Broglie wavelength for particles of species $a$ is
$\Lambda_{a} = 2 \pi \hbar / \sqrt{2m_{a} \pi kT}$.
In the classical description we get for the pressure, the compressibilty, 
and the coefficient of strain
\begin{equation} 
p_{\rm i}^{\rm id} = \frac{1}{k_{\rm T,i}^{\rm id}} 
                   =  \phi_{\rm S,i}^{\rm id} \; = \; 1 \;\; ,
\end{equation}
and for the internal energy and the isochoric specific heat 
\begin{equation}
u_{\rm i}^{\rm id} = c_{\rm V,i}^{\rm id} \; = \; \frac{3}{2} \;\; . 
\end{equation}
 
The ideal pressure of the electrons at any relativity and 
degeneracy (Chandrasekhar 1939, Cox \& Giuli 1968)
can be  calculated by 
\begin{equation} \label{Erelpress}
p_{\rm e}^{\rm id} = \frac{2}{n_{\rm e}\Lambda_{\rm e}^{3}}
      \left[ \; J_{3/2}(\psi,\lambda) +
      \frac{5}{4}\lambda J_{5/2}(\psi,\lambda) \; \right] \;\; ,
\end{equation}
with the correponding particle number density 
\begin{equation} \label{Ereldens}
n_{\rm e} = \frac{2}{\Lambda_{\rm e} ^{3}} \left[ \; J_{1/2}(\psi,\lambda) +
            \frac{3}{2} \lambda J_{3/2}(\psi,\lambda) \;\right] \;  \;\; .
\end{equation}  
The thermodynamical potentials of the electrons are characterized by the 
relativistic Fermi-Dirac integrals $J_{\nu} (\psi,\lambda)$, 
which depend on the degeneracy ($\psi$) and the relativity ($\lambda$)
parameters
\begin{equation} \label{Eplaspar}
\psi = \frac{\mu}{kT}  \;\;\; ,       \;\;\;\;\;\;\;\;\;\;\;
\lambda = \frac{kT}{m c^{2}}   \;\;\; .   
\end{equation}
Generally, the degeneracy parameter $\psi$ in (\ref{Eplaspar}) is
a function of the density and temperature defined by Eq.~(\ref{Ereldens}). 
In order to evaluate the free energy
\begin{equation} \label{Erelhelm}
f_{\rm e}^{\rm id} =  \psi - p_{\rm e}^{\rm id} 
\end{equation}
we have to determine $\psi$ explicitly from Eq.~(\ref{Ereldens}) by an 
inversion procedure which has  to be performed numerically.
Note, that $\psi$ in (\ref{Eplaspar}) is identical with the ideal 
contribution of $g$ in (\ref{Efdim1}), i.e. $\psi=g_{\rm e}^{\rm id}$.
Lamb (1974) and Lamb \& Van Horn (1975) evaluated the thermodynamical 
potentials applying the parametrizations of Eggleton \etal\ (1973) 
for the relativistic Fermi-Dirac integrals.
Johns \etal\ (1996) improved the accuracy of the polynomials given by
Eggleton \etal\ (1973).
Straniero (1988) calculated the complete set of thermodynamical functions
based on the expressions given in Eqs.~(\ref{Erelpress})-(\ref{Erelhelm})
by numerical integrations to determine the adiabatic temperature gradient.
Recently, Blinnikov \etal\ (1996) and Miralles \& Van Riper (1996) presented
parametrizations and used various approximations to evaluate 
the fully relativistic ideality for the set of thermodynamical potentials,
which are listed in Sect.~\ref{theomodel}. However, all calculational 
schemes are characterized by an immense effort in order to determine 
temperature- and density derivations even for asymptotic regions. 
For details, see e.g. Miralles \& Van Riper (1996). 
 
We pursue to include relativistic effects over a broad region
of astrophysically relevant densities and temperatures for
ideality and exchange.
Furthermore we evaluate the set of thermodynamical potentials analytically
avoiding the well-known noise problems of second-order quantities apparent
in purely numerical approaches.
\\
This can be realized by introducing two approximations: 
\\
{\bf a)} arbitrary degeneracy, but weak relativity ({\it case a}) and 
\\
{\bf b)} arbitrary relativity, but  strong degeneracy   ({\it case b})
\\
A numerical study on the density-temperature validity region
of these approximations was carried out in paper I. 
Note, that  {\it case a},
based on $\lambda$-expansions $\sim O(\lambda^{4})$,
is limited to $T \la 2 \cdot 10^{9} {\rm K}$ and
$\rho \la 10^6 {\rm g/cm^{3}}$, whereas  {\it case b},
based on $1/\psi$-Sommerfeld-Chandrasekhar expansions 
$\sim {\rm O}(\psi^{-6})$ holds for $\psi \ga 5$
(see Fig.\,1 in paper I).

Carrying out the expansions ($\sim O(\lambda^{4})$) in {\it case a}
we get for Eqs.~(\ref{Erelpress}) and (\ref{Ereldens})
\begin{equation}     \label{Eapide}
p_{\rm e}^{\rm id} = \frac{I_{3/2}(\psi)}{I_{1/2}(\psi)} \;
                     \frac{U^{\rm a}(\psi,\lambda)}{V^{\rm a}(\psi,\lambda)}
\end{equation}
\begin{equation}   \label{Ecane}
n_{\rm e} = \frac{2}{\Lambda_{\rm e}^{3}} I_{1/2}(\psi) \; 
                                        V^{\rm a}(\psi,\lambda) \;\; ,
\end{equation}
with the abbreviations $U^{\rm a}$ and $V^{\rm a}$ (not to be confused
with the internal energy or the volume)
\begin{equation}
U^{\rm a}  = 1 + \frac{15}{ 8} \lambda \frac{I_{5/2}}{I_{3/2}}
  \;  \left [1 + \frac{ 7}{16} \lambda \frac{I_{7/2}}{I_{5/2}}
  \;  \left (1 - \frac{ 3}{ 8} \lambda \frac{I_{9/2}}{I_{7/2}} \right) \right]
\end{equation}
and
\begin{equation}
V^{\rm a} =  1 + \frac{15}{ 8} \lambda \frac{I_{3/2}}{I_{1/2}}
  \;  \left [1 + \frac{ 7}{16} \lambda \frac{I_{5/2}}{I_{3/2}}
  \;  \left (1 - \frac{ 3}{ 8} \lambda \frac{I_{7/2}}{I_{5/2}} \right) \right]
\end{equation}   
taking into account relativistic corrections.   
$I_{\nu} = I_{\nu}(\psi) = \int_{0}^{\infty} dz \; z^{\nu}
({\rm e}^{z - \psi} + 1)^{-1}/\Gamma (\nu + 1)$ 
are the nonrelativistic Fermi-Dirac integrals considered
by parametrizations and expansions
(see e.g. paper I).

For the Helmholtz free energy in Eq.~(\ref{Erelhelm}) we have performed 
the inversion $\psi=\psi(n_{\rm e}, T)$ analytically (see paper I).
Using the Maxwell relation Eq.~(\ref{EUbas}) and 
Eqs.~(\ref{EPhi})-(\ref{EchiR}) we obtain for the internal energy, 
compressibility, and coefficient of strain
\begin{equation}  \label{Einna}
u_{\rm e}^{\rm id} = \frac{3}{2} p_{\rm e}^{\rm id} +      
                     \frac{15}{8} \lambda \frac{I_{5/2}}{I_{1/2}} \;
                     \frac{U^{\rm a}_{\lambda}}{V^{\rm a}} 
\end{equation}
\begin{equation}
\frac{1}{k_{\rm T,e}^{\rm id}} 
  = \frac{I_{1/2}}{I_{-1/2}} \; \frac{V^{\rm a}}{V^{\rm a}_{\psi}}
\end{equation}
\begin{equation}  \label{Estraina}
\phi_{\rm S,e}^{\rm id} = u_{\rm e}^{\rm id} + p_{\rm e}^{\rm id}
 -  \frac{3}{2} \; \frac{I_{1/2}}{I_{-1/2}} \;
    \left[ \frac{V^{\rm a}}{V^{\rm a}_{\psi}}
 +  \frac{5}{4} \lambda \frac{I_{3/2}}{I_{1/2}}  \;
    \frac{V^{\rm a}_{\lambda}}{V^{\rm a}_{\psi}} \right]  \;\; .
\end{equation}
Obviously, Eq.~(\ref{Einna}) results for  $\lambda \rightarrow 0$
in the well-known nonrelativistic relation
$u^{\rm id}=\frac{3}{2} p^{\rm id}$.
Calculating the derivatives of the free energy with respect to the
temperature in order to get the isochoric specific heat (Eq.~\ref{ECbas})
we have to execute explicitely
\begin{eqnarray}    \label{Ecvgena}
C_{\rm V} & = & - T \left[ \frac{\partial^2 F}{\partial T^2}  
                    + 2 \; \frac{{\rm d}\psi}{{\rm d}T} \;
                           \frac{\partial^2 F}{\partial \psi \partial T}
 \right. \nonumber \\         
& & \left.  \quad\quad\quad\quad\quad   
                    +      \left( \frac{{\rm d}\psi}{{\rm d}T}\right)^2 \;   
                           \frac{\partial^2 F}{\partial\psi^2}
                    +      \frac{{\rm d}^2\psi}{{\rm d}T^2} \;
               \frac{\partial F}{\partial\psi} \right]_{V,N_{a}} \; .  
\end{eqnarray}
The temperature derivatives of $\psi$ at constant
$V$ and $N_{\rm e}$ must be 
calculated from Eq.~(\ref{Ecane}) giving
\begin{equation}
T \frac{{\rm d}\psi}{{\rm d}T} = -  \frac{3}{2} \; 
                                    \frac{I_{1/2}}{I_{-1/2}} \;
                           \left[ \frac{V^{\rm a}}{V^{\rm a}_{\psi}}
 +              \frac{5}{4} \lambda \frac{I_{3/2}}{I_{1/2}}  \;
                                  \frac{V^{\rm a}_{\lambda}}{V^{\rm a}_{\psi}}
                            \right]  \;\; ,
\end{equation}
\begin{eqnarray}
T^{2}   \frac{{\rm d}^2\psi}{{\rm d}T^2} & = & 
          \frac{15}{4}
  \frac{I_{ 1/2}}{I_{-1/2}} \frac{V^{\rm a}}{V^{\rm a}_{\psi}}
- \left[\frac{3}{5}
  \frac{I_{ 1/2}}{I_{-1/2}} \frac{V^{\rm a}}{V^{\rm a}_{\psi}} \;
  \frac{I_{-3/2}}{I_{-1/2}} \frac{V^{\rm a}_{\psi\psi}}{V^{\rm a}_{\psi}}
        \right.  \nonumber \\
 & &    \left. 
+ \frac{3}{4} \lambda  \left( 
  \frac{I_{ 1/2}}{I_{-1/2}} \frac{V^{\rm a}_{\lambda\psi}}{V^{\rm a}_{\psi}}
- \frac{I_{ 3/2}}{I_{-1/2}} \frac{V^{\rm a}_{\lambda}}{V^{\rm a}_{\psi}} \;
  \frac{I_{-3/2}}{I_{-1/2}} \frac{V^{\rm a}_{\psi\psi}}{V^{\rm a}_{\psi}} 
                       \right)     
        \right.  \nonumber \\
 & &    \left.
+ \frac{15}{8} \lambda^{2}  
  \frac{I_{ 3/2}}{I_{-1/2}} \frac{V^{\rm a}_{\lambda}}{V^{\rm a}}
                       \left(
  \frac{I_{ 1/2}}{I_{-1/2}} \frac{V^{\rm a}_{\lambda\psi}}{V^{\rm a}_{\psi}}
        \right.  \right.  \nonumber \\
 & &    \left.   \left.
- \frac{1}{2}
  \frac{I_{ 3/2}}{I_{ 1/2}} \; \frac{V^{\rm a}_{\lambda}}{V^{\rm a}_{\psi}} 
  \frac{I_{-3/2}}{I_{-1/2}} \; \frac{V^{\rm a}_{\psi\psi}}{V^{\rm a}_{\psi}}
-         \frac{7}{30}   
  \frac{I_{ 5/2}}{I_{ 3/2}} \; \frac{V^{\rm a}_{\lambda\lambda}}
                                    {V^{\rm a}_{\lambda}}  \right) \right] .
\end{eqnarray}
Finally, Eq. (\ref{Ecvgena}) divided by $N_{\rm e}k$ 
yields the ideal part of the electronic specific heat 
\begin{eqnarray}   \label{ECVIida}
c_{\rm V,e}^{\rm id} & = & \frac{15}{4} \left[ \frac{I_{3/2}}{I_{1/2}} \;
     \frac{U^{\rm a}}{V^{\rm a}} - \frac{3}{5} \; \frac{I_{1/2}}{I_{-1/2}} \; 
     \frac{V^{\rm a}}{V^{\rm a}_{\psi}} \right]          \nonumber \\
 & & + \; \frac{75}{8} \lambda  \left[ \frac{I_{5/2}}{I_{1/2}} \;
          \frac{U^{\rm a}_{\lambda}}{V^{\rm a}}
     -    \frac{3}{5} \frac{I_{3/2}}{I_{-1/2}} \;
     \frac{V^{\rm a}_{\lambda}}{V^{\rm a}_{\psi}} \right]   \nonumber \\
 & & + \; \frac{105}{64} {\lambda}^{2} \left[ \frac{I_{7/2}}{I_{1/2}} \;
          \frac{U^{\rm a}_{\lambda\lambda}}{V^{\rm a}}
     -    \frac{15}{7} \; \frac{I_{3/2}}{I_{-1/2}} \;
     \frac{V^{\rm a}_{\lambda}}{V^{\rm a}_{\psi}} \; \frac{I_{3/2}}{I_{1/2}} \;
     \frac{V^{\rm a}_{\lambda}}{V^{\rm a}} \right]  ,
\end{eqnarray}
where the abbreviations are given by 
\begin{equation}
U^{\rm a}_{\lambda} = 1 + \frac{7}{8} \lambda \frac{I_{7/2}}{I_{5/2}}
  \; \left (1 - \frac{9}{16}    \lambda \frac{I_{9/2}}{I_{7/2}} \right)
\end{equation}
\begin{equation}
U^{\rm a}_{\lambda \lambda} = 1 - \frac{9}{8} \lambda \frac{I_{9/2}}{I_{7/2}}
\end{equation}   
\begin{equation}
V^{\rm a}_{\lambda} = 1 + \frac{7}{8} \lambda \frac{I_{5/2}}{I_{3/2}}  
  \; \left (1 - \frac{9}{16}    \lambda \frac{I_{7/2}}{I_{5/2}} \right)
\end{equation}  
\begin{equation}
V^{\rm a}_{\lambda \lambda} = 1 - \frac{9}{16} \lambda \frac{I_{7/2}}{I_{5/2}}
\end{equation}  
\begin{equation}
V^{\rm a}_{\psi} = 1 + \frac{15}{8}\lambda \frac{I_{1/2}}{I_{-1/2}}
  \; \left [1 + \frac{7}{16} \lambda \frac{I_{3/2}}{I_{1/2}}
  \; \left (1 - \frac{3}{8}  \lambda \frac{I_{5/2}}{I_{3/2}} \right) \right]
\end{equation}  
\begin{equation}
V^{\rm a}_{\psi \psi} = 1 + \frac{15}{8}\lambda \frac{I_{-1/2}}{I_{-3/2}}
  \; \left [1 + \frac{7}{16} \lambda \frac{I_{1/2}}{I_{1/2}}
  \; \left (1 - \frac{3}{8}  \lambda \frac{I_{3/2}}{I_{1/2}} \right) \right]
\end{equation}
\begin{equation}
V^{\rm a}_{\psi \lambda} = 1 + \frac{7}{8} \lambda \frac{I_{3/2}}{I{1/2}}
  \; \left (1 - \frac{9}{16} \lambda \frac{I_{5/2}}{I_{3/2}} \right) \;\; .
\end{equation}
From these relations we can read off immediately various asymptotics.
The nonrelativistic ($\lambda \ll 1$) and nondegenerate limit ($\psi \ll -1$)
of Eq.~(\ref{ECVIida}) results to the well-known classical value 
\begin{equation}
c_{\rm V,e}^{\rm id}  =  \frac{3}{2} \;\; .
\end{equation}
Furthermore in the nonrelativistic limit and at arbitrary degeneracy 
the comparison of Eqs.~(\ref{Estraina}) and (\ref{ECVIida}) yields
\begin{equation}
\phi_{\rm S,e}^{\rm id} = \frac{2}{3} c_{\rm V,e}^{\rm id}  \;\; .
\end{equation}
{\it Case b} describes the regime where the
Sommerfeld-Chandrasekhar expansion (Chandrasekhar 1939, Cox \& Giuli 1968)
becomes valid.
The degeneracy parameter $\psi$ is related for the strong-degenerate limit by
\begin{equation}  \label{Ealphzo}
\psi \lambda =  \sqrt{ 1 + \alpha^{2}} - 1   \;\;  ,
\end{equation}
where $\alpha = p_{\rm F}/mc$ is the fraction of the 
Fermi- to the relativistic momentum. 
The Sommerfeld-Chandrasekhar expansion with $\sim {\rm O}(\psi^{-6})$ provides
(Cox \& Giuli 1968)
\begin{equation}
p_{\rm e}^{\rm id} = \frac{1}{4\lambda} \; \frac{U^{\rm b}(\psi,\lambda)}
                                                {V^{\rm b}(\psi,\lambda)} \;\;,
\end{equation}
\begin{equation} \label{Ecbne}
n_{\rm e} = \frac{2}{\Lambda_{\rm e}^{3}} \sqrt{\frac{2}{\pi}}
            \frac{\alpha^{3}}{3 \lambda^{3/2}}  \; V^{\rm b}(\psi,\lambda) 
          = \left(\frac{m c}{\hbar}\right)^{3} \frac{\alpha^{3}}{3 \pi^{2}}
            \; V^{\rm b}(\psi,\lambda)  \; ,
\end{equation}
with the abbreviations 
\begin{eqnarray}
U^{\rm b} & = & \sqrt{1 + \alpha^{2}} \left[ 1 - \frac{3}{2 \alpha^{2}}
  + \frac{3}{2 \alpha^{3} \sqrt{1 + \alpha^{2}}}
    \ln\left(\alpha+\sqrt{1+\alpha^{2}}\right) 
\right. \nonumber \\
 & & \left. \quad\quad\quad\quad\quad 
     + \frac{2 \pi^{2} \lambda^{2}}{\alpha^{2}} 
       \left(1 - \frac{7}{60} \frac{\pi^{2} \lambda^{2}}{\alpha^{4}}
       \left(1 - 2 \alpha^{2} \right) \right) \right]  
\end{eqnarray}
and
\begin{equation}
V^{\rm b} =  1 + \frac{\pi^{2} \lambda^{2}}{2 \alpha^{4}} \left( 
  1 + 2 \alpha^{2}+\frac{7}{20}\frac{\pi^{2}\lambda^{2}}{\alpha^{4}} \right)
   \;\; .
\end{equation}
For the Helmholtz free energy represented by Eq.~(\ref{Erelhelm}) 
the inversion $\psi=\psi(n_{\rm e}, T)$ for {\it case b} is already given 
by Eq.~(\ref{Ealphzo}).
In order to correct Eq.~(\ref{Ealphzo}) to the order $\sim {\rm O}(\psi^{-6})$
analytical inversions have been performed by Yakovlev \& Shalybkov (1989)
and paper I.
 
For the internal energy, compressibility, and coefficient of strain we get
\begin{equation}  \label{Einnb}
u_{\rm e}^{\rm id}  =  
\frac{1}{\lambda} \left[  \sqrt{1 + \alpha^{2}} - 1 +
       \frac{1}{4} \left( \frac{U^{\rm b}}{V^{\rm b}} +
                          \frac{U^{\rm b}_{\lambda}}{V^{\rm b}}\right) \right]
\end{equation}
\begin{equation}
\frac{1}{k_{\rm e}^{\rm id}} = \frac{1}{\lambda} \; 
 \left[ \frac{\alpha^{2}}{\sqrt{1 + \alpha^{2}}} 
 \left( \frac{V^{\rm b}}{3 V^{\rm b}+V^{\rm b}_{\alpha}} \right) \right] 
\end{equation}
\begin{equation}
\phi_{\rm e}^{\rm id} = \frac{1}{\lambda} \;
      \left[ \; \frac{1}{4} \frac{U^{\rm b}_{\lambda}}{V^{\rm b}}
               -  \frac{\alpha^{2}}{\sqrt{1 + \alpha^{2}}}
  \left( \frac{V^{\rm b}_{\lambda}}{3 V^{\rm b}+V^{\rm b}_{\alpha}} \right)
         \right]  \;\; .
\end{equation}
Neglecting the order $\psi^{-6}$ in Eq.~(\ref{Einnb}) we get the 
Chandrasekhar approximation for the internal energy  
(Chandrasekhar 1939, Cox \& Giuli 1968, Lamb 1974, 
Eliezer \etal\ 1986, Yakovlev \& Shalybkov 1989)
\begin{eqnarray}    \label{Einnbexpl} 
u_{\rm e}^{\rm id} & = & \frac{1}{N_{\rm e} k T} \; \frac{m c^{2}}{8 \pi^{2}} 
                  \left( \frac{m c}{\hbar}\right)^3 \;\; \times 
 \nonumber \\ 
 & &  \left[
    \alpha \sqrt{1 + \alpha^{2}} ( 1 + 2 \alpha^{2}) - \frac{8}{3 \alpha^{3}}
                 -\ln\left(\alpha+\sqrt{1+\alpha^{2}}\right)
\right. \nonumber \\
 & & \left.  \quad
     + \frac{4}{3} \frac {\pi^{2} \lambda^{2}}{\alpha}
    \left( \sqrt{1 + \alpha^{2}} (1 + 3 \alpha^{2})-(1 + 2 \alpha^{2}) \right)
    \right] \; .
\end{eqnarray}
The specific heat Eq.~(\ref{ECbas}) can be calculated for {\it case b} via
\begin{eqnarray}    \label{Ecvgenb}
C_{\rm V} & = & - T \left[ \frac{\partial^2 F}{\partial T^2}
                    + 2 \; \frac{{\rm d}\alpha}{{\rm d}T} \;
                           \frac{\partial^2 F}{\partial \alpha \partial T}
 \right. \nonumber \\
& & \left.  \quad\quad\quad\quad\quad
                    +      \left( \frac{{\rm d}\alpha}{{\rm d}T}\right)^2 \;
                           \frac{\partial^2 F}{\partial\alpha^2}
                    +      \frac{{\rm d}^2\alpha}{{\rm d}T^2} \;
               \frac{\partial F}{\partial\alpha} \right]_{V,N_{a}} \;\; .
\end{eqnarray}
The temperature derivatives of $\alpha$ at constant $V$ and $N_{\rm e}$ must 
be calculated from Eq.~(\ref{Ecbne}) 
\begin{equation} \label{Edalpdt}
\frac{T}{\alpha} \; \frac{{\rm d}\alpha}{{\rm d}T} =
     - \frac{ V^{\rm b}_{\lambda}}{3 V^{\rm b} + V^{\rm b}_{\alpha}}  \;\; , 
\end{equation}
\begin{eqnarray} \label{Edalp2dt}
\frac{T^{2}}{\alpha} \; \frac{{\rm d}^2\alpha}{{\rm d}T^2} & = &
             \frac{T}{\alpha} \frac{{\rm d}\alpha}{{\rm d}T}
      \left[ \frac{V^{\rm b}_{\lambda\lambda}}{V^{\rm b}_{\lambda}}
    +    2   \frac{T}{\alpha} \frac{{\rm d}\alpha}{{\rm d}T}   
             \frac{ V^{\rm b}_{\alpha\lambda}}{V^{\rm b}_{\lambda}}
        \right.  \nonumber \\
 &  &   \left. \quad\quad\quad\quad\quad
    -  \left( \frac{T}{\alpha} \frac{{\rm d}\alpha}{{\rm d}T} \right)^{2}
       \left( \frac{ 12 V^{\rm b}}{V^{\rm b}_{\lambda}} 
            - \frac{ V^{\rm b}_{\alpha\alpha}}{V^{\rm b}_{\lambda}} \right)
      \right] \; .      
\end{eqnarray}
Finally, Eq.~(\ref{Ecvgenb}) divided by $N_{\rm e}k$ yields 
\begin{eqnarray} \label{ECVIidb}
c_{\rm V,e}^{\rm id} & = & - \frac{1}{\lambda}  
  \left\{ \left(\sqrt{1+\alpha^{2}}-1\right)
                    \frac{V^{\rm b}_{\lambda\lambda}}{V^{\rm b}}   
      - \frac{1}{4} \frac{U^{\rm b}_{\lambda\lambda}}{V^{\rm b}}
        \right.  \nonumber \\
 &  &   \left.
 + 2\frac{T}{\alpha} \frac{{\rm d}\alpha}{{\rm d}T}
 \left[ \left(\sqrt{1+\alpha^{2}}-1\right)
        \left(\frac{3 V^{\rm b}_{\lambda}}{V^{\rm b}} 
  +           \frac{  V^{\rm b}_{\alpha \lambda}}{V^{\rm b}}\right) \right]
        \right.  \nonumber \\
 &  &   \left.
  - \left( \frac{T}{\alpha} \frac{{\rm d}\alpha}{{\rm d}T} \right)^{2}
 \left[ 6   \left( \sqrt{1+\alpha^{2}}-1 \right) 
 \left( 1 +  \frac{       V^{\rm b}_{\alpha}}{V^{\rm b}}
  +     \frac{1}{6} \frac{V^{\rm b}_{\alpha\alpha}}{V^{\rm b}} \right) 
        \right. \right.  \nonumber \\
 &  &   \left.  \left. 
  -                      \frac{3U^{\rm b}}{V^{\rm b}} 
  +          \frac{1}{4} \frac{U^{\rm b}_{\alpha\alpha}}{V^{\rm b}}
  -          \frac{\alpha^{2}}{(1+\alpha^{2})^{3/2}}
 \left(1 + 2 (1+\alpha^{2}) \frac{V^{\rm b}_{\alpha}}{V^{\rm b}} \right)\right]
        \right.   \nonumber \\
 &  &   \left.  
 + \frac{T^{2}}{\alpha} \frac{{\rm d}^2\alpha}{{\rm d}T^2}
 \left[ \left( \sqrt{1+\alpha^{2}}-1 \right)
        \left( 3 + \frac{V^{\rm b}_{\alpha}}{V^{\rm b}} \right)
        \right] \right\}  \;\; . 
\end{eqnarray}
Here the abbreviations labeled by the superscript b  
are given by the simple derivatives with $X=\{U^{\rm b},V^{\rm b}\}$
\begin{equation}
X_{\lambda} = \lambda \frac{\partial}{\partial \lambda}  X \;\; , \quad
X_{\lambda\lambda} = \lambda^2 \frac{\partial^2}{\partial \lambda^2}  X \;\; ,
\end{equation}
\begin{equation} 
X_{\alpha} = \alpha \frac{\partial}{\partial \alpha}  X \;\; , \quad
X_{\alpha\alpha} = \alpha^2 \frac{\partial^2}{\partial \alpha^2}  X \;\; ,
\end{equation}
\begin{equation}
X_{\alpha\lambda} = \alpha \lambda \frac{\partial^2}
                       {\partial \alpha \partial \lambda}  X \;\; .
\end{equation}
Eq.~(\ref{ECVIidb}) yields by neglecting the order $\psi^{-6}$
(Chandrasekhar 1939, Eliezer \etal\ 1986, Yakolev \& Shalybkov 1989)
\begin{equation}
c_{\rm V,e}^{\rm id} = \frac{\pi^2 \lambda}{\alpha^{2}} 
                                   \sqrt{1+ \alpha^{2}} \;\; ,
\end{equation}
which delivers the strong-relativistic ($\alpha \gg 1$)  
\begin{equation}
c_{\rm V,e}^{\rm id}  =  \frac{\pi^2}{\psi}  
\end{equation}
and the nonrelativistic ($\alpha \ll 1$) 
\begin{equation}
c_{\rm V,e}^{\rm id}  =  \frac{\pi^2}{2 \psi}    
\end{equation}
limiting laws.
Let us remark that both Eqs.~(\ref{ECVIida}) and (\ref{ECVIidb}) 
have in the degenerate, weak-relativistic limit a wide overlap, which we use to
combine {\it case a} and {\it b} in order to realize numerical continuity 
(see paper I).
\subsection{Exchange}
The lowest order exchange (Hartree-Fock) free energy for relativistic 
electrons is given by 
\begin{equation} \label{Efeex}
f_{\rm ee}^{\rm x} = \frac{e^{2}}{\pi \Lambda_{\rm e}^{4}}
   \frac{1}{n_{\rm e}kT} \, J_{\rm rel}^{\rm x}(\psi, \lambda) \;\; ,
\end{equation}   
where $J_{\rm rel}^{\rm x}(\psi, \lambda)$ is the  relativistic Hartree-Fock
integral (Kovetz \etal\ 1972, paper I). 

The $\lambda$-expansion in {\it case a} results in   
\begin{equation}
f_{\rm ee}^{\rm x} = -   \frac{e^{2}}{kT \Lambda_{\rm e}} \;
                         \frac{I^{\rm x}(\psi)}{I_{1/2}(\psi)} \; 
  \frac{W^{\rm a}(\psi,\lambda)}{V^{\rm a}(\psi,\lambda)} \;\; ,
\end{equation}
where
\begin{eqnarray}
W^{\rm a} & = & 1 - \frac{3}{8} \; \lambda \; 
                    \frac{I_{1/2}^{2}}{I^{\rm x}} \;
        \left [1 + \frac{9}{8}   \lambda \frac{I_{3/2}}{I_{1/2}}
        \left(1 + \frac{11}{72} \lambda \frac{I_{3/2}}{I_{1/2}} \; \times
     \right.\right.   \nonumber \\
 & & \left. \left. \quad\quad\quad\quad\quad\quad\quad\quad\quad
        \left(1 - \frac{25}{11} \lambda \frac{I_{5/2}}{I_{3/2}}
                  \frac{I_{1/2}}{I_{3/2}} \right) \right) \right] \;.
\end{eqnarray}
Parametrizations for the nonrelativistic Hartree-Fock integral 
\begin{equation}
I^{\rm x}(\psi)=
\int_{-\infty}^{\psi} d \psi ^{\prime} \, I_{-1/2}^{2}(\psi^{\prime})
\end{equation}
are given by Perrot \& Dharma-wardana (1984) and Kalitkin \& Ritus (1986)
(see Stolzmann \& Bl\"ocker 1996b).
Note, that for $\lambda=0$ the following relation holds 
\begin{equation}
J_{\rm rel}^{\rm x}(\psi, \lambda = 0) \; = \; -2 \pi \; I^{\rm x}(\psi)  \;\; .
\end{equation}
The exchange pressure   
$p_{\rm ee}^{\rm x} = g_{\rm ee}^{\rm x} - f_{\rm ee}^{\rm x}$
is determined via the exchange Gibbs energy (cf.\ paper I)
\begin{equation}
g_{\rm ee}^{\rm x}  =  - \frac{e^{2}}{kT \Lambda_{\rm e}} \; I_{-1/2} \;
                         \frac{W^{\rm a}_{\psi}}{V^{\rm a}_{\psi}} \;\; ,
\end{equation} 
where
\begin{eqnarray}
W^{\rm a}_{\psi} & = & 1 - \frac{3}{4} \lambda \frac{I_{1/2}}{I_{-1/2}}
          \left [1 + \frac{9}{16}  \lambda \frac{I_{1/2}}{I_{-1/2}}
          \left (1 + \frac{I_{3/2}}{I_{1/2}}\frac{I_{-1/2}}{I_{1/2}}
 \right. \right. \nonumber \\
 & &  \left. \left.  \quad\quad\quad\quad\quad
 - \frac{1}{24} \lambda \frac{I_{3/2}}{I_{1/2}}  
           \left (1 + \frac{25}{3} \frac{I_{5/2}}{I_{3/2}}
          \frac{I_{-1/2}}{I_{1/2}} \right) \right) \right] 
\end{eqnarray}
For the exchange contribution of the internal energy, compressibility,
and coefficient of strain we get 
\begin{equation}
u_{\rm ee}^{\rm x} = \frac{3}{2} p_{\rm ee}^{\rm x} 
   - \frac{1}{2} \;\frac{e^{2}}{kT \Lambda_{\rm e}}
     \frac{I^{\rm x}}{I_{1/2}} \;
     \frac{W^{\rm a}_{\lambda}}{V^{\rm a}}
   + \frac{15}{8} \lambda \frac{I_{3/2}}{I_{1/2}} \;
     \frac{V^{\rm a}_{\lambda}}{V^{\rm a}} \; g_{\rm ee}^{\rm x} 
\end{equation}
\begin{equation}
\frac{1}{k_{\rm T,ee}^{\rm x}} = - \frac{e^{2}}{kT \Lambda_{\rm e}} \;
          \frac{I_{1/2} I_{-3/2}}{I_{-1/2}} \;
          \frac{V^{\rm a}}{V^{\rm a}_{\psi}} \;
 \left[2 \; \frac{W^{\rm a}_{\psi \psi}}{V^{\rm a}_{\psi}} -
     \frac{V^{\rm a}_{\psi \psi}}{V^{\rm a}_{\psi}} \;
     \frac{W^{\rm a}_{\psi}}{V^{\rm a}_{\psi}} \right] 
\end{equation}
\begin{eqnarray}
\phi_{\rm S,ee}^{\rm x} & = & p_{\rm ee}^{\rm x} + u_{\rm ee}^{\rm x} 
           + \frac{1}{2} \; \frac{e^{2}}{kT \Lambda_{\rm e}} \; I_{-1/2} \;
\left[ 3 \; \frac{I_{1/2}}{I_{-1/2}}\frac{I_{-3/2}}{I_{-1/2}} \;\; \times
        \right. \nonumber \\
 & &    \left.            
\left(2 \frac{W^{\rm a}_{\psi \psi}}{V^{\rm a}_{\psi}}
           -  \frac{V^{\rm a}_{\psi \psi}}{V^{\rm a}_{\psi}}
              \frac{W^{\rm a}_{\psi}}{V^{\rm a}_{\psi}} \right)
      \left(  \frac{V^{\rm a}}{V^{\rm a}_{\psi}} +
              \frac{5}{4} \lambda \; \frac{I_{3/2}}{I_{1/2}}
              \frac{V^{\rm a}_{\lambda}}{V^{\rm a}_{\psi}} \right)
        \right. \nonumber \\
 & &    \left.  \quad\quad\quad\quad\quad
 + \frac{W^{\rm a}_{\psi \lambda}}{V^{\rm a}_{\psi}}
 + \frac{15}{4} \lambda \; \frac{I_{1/2}}{I_{-1/2}}
    \frac{V^{\rm a}_{\psi \lambda}}{V^{\rm a}_{\psi}} \;
    \frac{W^{\rm a}_{\psi}}{V^{\rm a}_{\psi}} \right] \; .
\end{eqnarray}
Using (\ref{Ecvgena}) we obtain for the exchange part of the isochoric 
specific heat 
\begin{eqnarray}
c_{\rm V,e}^{\rm x} & = & \frac{e^{2}}{kT \Lambda_{\rm e}} \; 
   \left[ \frac{15}{4}
   \left[\frac{I^{\rm x}}{I_{1/2}} \; 
       \left( \frac{W^{\rm a}}{V^{\rm a}}
- \frac{2}{3} \frac{W^{\rm a}_{\lambda}}{V^{\rm a}} 
+ \frac{1}{5} \frac{W^{\rm a}_{\lambda\lambda}}{V^{\rm a}} \right)
        \right.  \right.   \nonumber \\
 &  &   \left.   \left. \quad\quad\quad 
+ \frac{3}{5} \; \frac{I_{1/2}I_{-3/2}}{I_{-1/2}} \; 
                 \frac{V^{\rm a}}{V^{\rm a}_{\psi}}   
    \left( 2 \;  \frac{W^{\rm a}_{\psi\psi}}{V^{\rm a}_{\psi}}
  -              \frac{V^{\rm a}_{\psi\psi}}{V^{\rm a}_{\psi}}   
                 \frac{W^{\rm a}_{\psi}}{V^{\rm a}} \right)
        \right.  \right. \nonumber \\
 &  &   \left.   \left.  \quad\quad\quad\quad\quad\quad\quad\quad\quad
- I_{-1/2} \left( \frac{W^{\rm a}_{\psi}}{V^{\rm a}_{\psi}} - \frac{2}{5}
           \frac{W^{\rm a}_{\psi\lambda}}{V^{\rm a}_{\psi}} \right)\right]
      \right. \nonumber \\
 & &  \left.
  + \frac{75}{8} \; \lambda
   \left[-\frac{I_{3/2}I_{-1/2}}{I_{1/2}} \;
          \frac{V^{\rm a}_{\lambda}}{V^{\rm a}} \;
   \left( \frac{W^{\rm a}_{\psi}}{V^{\rm a}_{\psi}} - \frac{1}{5} \;
          \frac{W^{\rm a}_{\psi\lambda}}{V^{\rm a}_{\psi}} \right)
      \right. \right. \nonumber \\
 & &  \left. \left. \quad\quad
  + \frac{3}{5} \; \frac{I_{3/2}I_{-3/2}}{I_{1/2}} \;
    \frac{V^{\rm a}_{\lambda}}{V^{\rm a}_{\psi}} \;
    \left( 2 \frac{W^{\rm a}_{\psi\psi}}{V^{\rm a}_{\psi}} -
             \frac{V^{\rm a}_{\psi\psi}}{V^{\rm a}_{\psi}} \;
             \frac{W^{\rm a}_{\psi}}{V^{\rm a}_{\psi}} \right)
      \right. \right. \nonumber \\
 & &  \left. \left.  \quad\quad\quad\quad\quad\quad\quad\quad\quad
                     \quad\quad\quad
  + \frac{3}{5} \; I_{1/2} \;
             \frac{V^{\rm a}_{\psi\lambda}}{V^{\rm a}_{\psi}} \;
             \frac{W^{\rm a}_{\psi}}{V^{\rm a}_{\psi}}  \right]
      \right. \nonumber \\
 & &  \left.
 + \frac{225}{64} \lambda^{2}
      \left[ \frac{I_{3/2}^{2}I_{-3/2}}{I_{1/2}I_{-1/2}}
             \frac{V^{\rm a}_{\lambda}}{V^{\rm a}{\psi}}
     \left(2 \frac{W^{\rm a}_{\psi\psi}}{V^{\rm a}}
 -           \frac{V^{\rm a}_{\psi\psi}}{V^{\rm a}_{\psi}}
             \frac{W^{\rm a}_{\psi}}{V^{\rm a}} \right)
      \right. \right. \nonumber \\
 & &  \left.  \left.
  + \; 4 \frac{I_{3/2}I_{-3/2}}{I_{1/2}} \left( \frac{I_{1/2}}{I_{-1/2}}
         \frac{V^{\rm a}_{\lambda}}{V^{\rm a}_{\psi}}
         \frac{V^{\rm a}_{\psi\lambda}}{V^{\rm a}_{\psi}}
 - \frac{7}{30} \frac{I_{5/2}}{I_{3/2}}
         \frac{V^{\rm a}_{\lambda\lambda}}{V^{\rm a}_{\psi}} \right)
                                                     \right] \right]
\end{eqnarray}
with the relativistic corrections 
\begin{eqnarray}  
W^{\rm a}_{\lambda} & = & 1 + \frac{3}{8} \lambda \; 
                              \frac{I_{1/2}^{2}}{I^{\rm x}} \;
\left [1 + \frac{27}{8}   \lambda \frac{I_{3/2}}{I_{1/2}}
\left (1 + \frac{55}{216} \lambda \frac{I_{3/2}}{I_{1/2}} \; \times
      \right. \right. \nonumber \\
 & &  \left.  \left.   \quad\quad\quad\quad\quad\quad\quad\quad\quad\quad
\left (1 - \frac{25}{11}  \frac{I_{5/2}}{I_{3/2}}
           \frac{I_{1/2}}{I_{3/2}} \right) \right) \right]
\end{eqnarray}
\begin{eqnarray}
W^{\rm a}_{\lambda \lambda} & = & 1 + \frac{1}{8} \lambda \; 
                                      \frac{I_{1/2}^{2}}{I^{\rm x}} \;
\left [1 - \frac{27}{8}   \lambda \frac{I_{3/2}}{I_{1/2}}
\left (1 + \frac{55}{72}  \lambda \frac{I_{3/2}}{I_{1/2}} \; \times
      \right. \right. \nonumber \\
 & &  \left.  \left.   \quad\quad\quad\quad\quad\quad\quad\quad\quad\quad
\left (1 - \frac{25}{11}  \frac{I_{5/2}}{I_{3/2}}
           \frac{I_{1/2}}{I_{3/2}} \right) \right) \right]
\end{eqnarray}
\begin{eqnarray}
W^{\rm a}_{\psi \psi} & = & 1 - \frac{3}{8} \lambda 
                     \frac{I_{ 1/2}}{I_{-1/2}}
          \left [1 + \frac{I_{-1/2}}{I_{-3/2}}
                     \frac{I_{-1/2}}{I_{ 1/2}} 
      \right.  \nonumber \\
 & &  \left.  \quad 
 + \frac{9}{16} \lambda \frac{I_{ 3/2}}{I_{ 1/2}}
           \left( 1 + 3 \frac{I_{-1/2}}{I_{-3/2}}
                        \frac{I_{ 1/2}}{I_{ 3/2}}
 - \frac{7}{9} \lambda  \frac{I_{-1/2}}{I_{-3/2}}
      \right. \right. \times  \nonumber \\        
 & &  \left.  \left. \quad\quad
          \left (1 + \frac{3}{28} \frac{I_{ 1/2}}{I_{-1/2}}
                                  \frac{I_{ 1/2}}{I_{ 3/2}}
                  + \frac{25}{28} \frac{I_{ 5/2}}{I_{ 3/2}}
                                  \frac{I_{-3/2}}{I_{-1/2}}
                                  \right) \right) \right]
\end{eqnarray}
\begin{eqnarray}
W^{\rm a}_{\psi \lambda} & = & 1 + \frac{3}{4} \lambda \;
                                 \frac{I_{ 1/2}}{I_{-1/2}}
\left [1 + \frac{27}{16} \lambda \frac{I_{ 3/2}}{I_{ 1/2}}
\left (1 +                       \frac{I_{ 1/2}}{I_{-1/2}} 
                                 \frac{I_{ 1/2}}{I_{ 3/2}}
     \right. \right.  \nonumber \\
& &  \left.  \left.  \quad\quad\quad\quad\quad\quad
         - \frac{125}{216}\lambda\frac{I_{ 5/2}}{I_{ 3/2}}
\left (1 + \frac{3}{25}          \frac{I_{ 3/2}}{I_{ 5/2}} 
                                 \right) \right) \right] \; .
\end{eqnarray}
The Sommerfeld-Chandrasekhar expansion for the exchange free energy in
{\it case b}
(Salpeter \& Zapolsky 1967, Kovetz \etal\ 1972, Lamb 1974)
delivers for $\sim {\rm O}(\psi^{-6})$ (see paper I)
\begin{equation}   \label{Efxb}
f_{\rm ee}^{\rm x} = - \frac{e^{2}}{kT \Lambda_{\rm e}} \;
                       \frac{3}{2 \alpha^{3}} \;
                       \frac{1}{\sqrt{2 \pi \lambda}} \;
                       \frac{W^{\rm b}(\alpha,\lambda)}
                            {V^{\rm b}(\alpha,\lambda)}  \;\; ,
\end{equation}  
where
\begin{eqnarray}
W^{\rm b} & = & \frac{3 B^{2}}{2(1+\alpha^{2})}
    - 3 \alpha B + \frac{3}{2} \alpha^{2} + \frac{1}{2} \alpha^{4}
 \nonumber \\
 & &  + \frac{\pi^{2} \lambda^{2}}{3} \left[ -0.7046
  + 2 \ln \frac{2 \alpha^{2}}{\lambda} + \alpha^{2}-\frac{3 B}{\alpha} \right]
 \nonumber \\
 & &  + \frac{\pi^{4} \lambda^{4}}{18} \left[ 1 - \frac{11}{10 \alpha^{2}}
    \left( 1 + \frac{37}{11 \alpha^{2}} + \frac{63 B}{11 \alpha^{3}}
 \right) \right]
\end{eqnarray}
with $B=\sqrt{1+\alpha^{2}}\ln\left(\alpha+\sqrt{1+\alpha^{2}}\right)$.
\\
With the Gibbs energy (see paper I)
\begin{equation}
g_{\rm ee}^{\rm x}  = - \frac{e^{2}}{kT \Lambda_{\rm e}} \;
                        \frac{3}{2 \alpha^{3}} \;
                        \frac{1}{\sqrt{2 \pi \lambda}} \;
                        \frac{W^{\rm b}_{\alpha}}
                           {3 V^{\rm b} + V^{\rm b}_{\alpha}}
\end{equation}
it is possible to provide the exchange pressure 
$p_{\rm ee}^{\rm x} = g_{\rm ee}^{\rm x} - f_{\rm ee}^{\rm x}$ again.

\begin{figure}
\centering
\epsfxsize=0.49\textwidth
\mbox{\epsffile{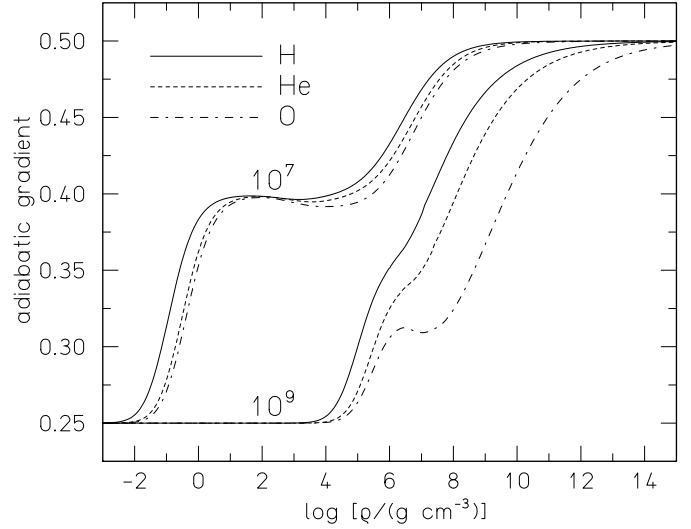}}
\caption[nab_H_HE_O_T79]
{Adiabatic temperature gradient $\nabla_{\rm ad}$ vs.\ density at $T=10^{7}{\rm K}$   
 and $T=10^{9}{\rm K}$ for various elements.
 The curves refer to the sum of  ideality, exchange and radiation.
 The presentation towards extremely high densities serves only
 to illustrate the asymptotic behaviour of the adiabatic gradient in our model
 (processes relevant to very high densities as, e.g., pycnonuclear reactions 
 or electron captures, are not considered).
}                           \label{Fnab_H_HE_O_ixr_T79}
\end{figure}

\begin{figure}
\centering
\epsfxsize=0.49\textwidth
\mbox{\epsffile{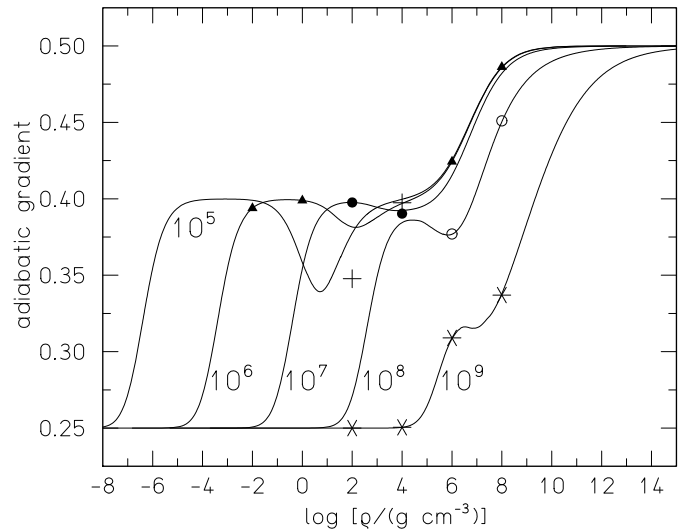}}
\caption[nab_C_ixr_T5_9]
{Adiabatic temperature gradient $\nabla_{\rm ad}$ vs.\ density for carbon along various
 isotherms compared with Lamb (1974)
 ($+ :        10^{5}{\rm K}$,
  $\triangle: 10^{6}{\rm K}$,
  $\bullet:   10^{7}{\rm K}$,
  $\circ:     10^{8}{\rm K}$,
  $\ast  :    10^{9}{\rm K}$).
The curves refer to the sum of ideality, exchange and radiation.
 The presentation towards extremely high densities serves only      
 to illustrate the asymptotic behaviour of the adiabatic gradient in our model
 (processes relevant to very high densities as, e.g., pycnonuclear reactions
 or electron captures, are not considered).
}                                      \label{Fnab_C_ixr_T5_9}
\end{figure}

In a straightforward manner,
the exchange contribution for internal energy, compressibility,
and coefficient of strain are represented by 
\begin{eqnarray}
u_{\rm ee}^{\rm x} & = & \frac{e^{2}}{kT \Lambda_{\rm e}} \;
                         \frac{3}{2 \alpha^{3}} \;
                         \frac{1}{\sqrt{2 \pi \lambda}} \;
 \left[\frac{W^{\rm b}_{\lambda}}{V^{\rm b}}
    -  \frac{W^{\rm b}_{\lambda}}{V^{\rm b}}
       \right. \nonumber \\
 & &   \left.  \quad\quad\quad\quad\quad\quad\quad\quad\quad\quad
    +  \; \frac{W^{\rm b}_{\alpha }}{V^{\rm b}}
\left(\frac{V^{\rm b}_{\lambda}}{3 V^{\rm b}+V^{\rm b}_{\alpha}}\right)\right]
\end{eqnarray}
\begin{eqnarray}
\frac{1}{k_{\rm T,ee}^{\rm x}} & = & \frac{e^{2}}{kT \Lambda_{\rm e}}
   \frac{3}{2 \alpha^{3}} \frac{1}{\sqrt{2 \pi \lambda}}
   \left( \frac{V^{\rm b}}{3 V^{\rm b} + V^{\rm b}_{\alpha}} \right)^{2}
  \times    \nonumber \\
 & &   \quad\quad  \left[\frac{W^{\rm b}_{\alpha\alpha}}{V^{\rm b}}
 - \frac{6 W^{\rm b}_{\alpha}}{3 V^{\rm b} + V^{\rm b}_{\alpha}}
   \left(1+\frac{V^{\rm b}_{\alpha}}{V^{\rm b}} + \frac{1}{6}
           \frac{V^{\rm b}_{\alpha\alpha}}{V^{\rm b}} \right)  \right] 
\end{eqnarray}
\begin{eqnarray}
\phi_{\rm S,ee}^{\rm x} & = & p_{\rm ee}^{\rm x} + u_{\rm ee}^{\rm x}
  -   g_{\rm ee}^{\rm x} \left[ 1 +
    \frac{3 V^{\rm b}_{\lambda} + V^{\rm b}_{\alpha\lambda}}
         {3 V^{\rm b} + V^{\rm b}_{\alpha}}
  - \frac{W^{\rm b}_{\alpha\lambda}}{W^{\rm b}_{\alpha}} \right. \nonumber \\
 & & \left.  \quad\quad
  - \frac{V^{\rm b}_{\lambda}}{3 V^{\rm b}+ V^{\rm b}_{\alpha}}
    \left(2 + \frac{4 V^{\rm b}_{\lambda} + V^{\rm b}_{\alpha\alpha}}
         {3 V^{\rm b} + V^{\rm b}_{\alpha}}
 - \frac{W^{\rm b}_{\alpha\lambda}}{W^{\rm b}_{\alpha}} \right) \right] \; .
\end{eqnarray}  
The exchange specific heat can be formulated by 
\begin{eqnarray}  \label{Ecvxb}
c_{\rm V,e}^{\rm x} & = & - \frac{e^{2}}{kT \Lambda_{\rm e}} \;
    \frac{3}{2 \alpha^{3}} \; \frac{1}{\sqrt{2 \pi \lambda}} 
\left[ \frac{W^{\rm b}_{\lambda\lambda}}{V^{\rm b}}
 + 2\frac{T}{\alpha} \frac{{\rm d}\alpha}{{\rm d}T}
   \frac{W^{\rm b}_{\alpha\lambda}}{V^{\rm b}}
        \right.  \nonumber \\
 &  &   \left.  \quad\quad\quad\quad\quad\quad
 + \left( \frac{T}{\alpha} \frac{{\rm d}\alpha}{{\rm d}T} \right)^{2}
   \frac{W^{\rm b}_{\alpha\alpha}}{V^{\rm b}}
 + \frac{T^{2}}{\alpha} \frac{{\rm d}^2\alpha}{{\rm d}T^2}
   \frac{W^{\rm b}_{\alpha}}{V^{\rm b}}
\right] 
\end{eqnarray}
The temperature derivatives of $\alpha$ in Eq.~(\ref{Ecvxb}) are given 
by (\ref{Edalpdt}) and (\ref{Edalp2dt}).
The derivatives of $W$ with respect to $\lambda$ and $\alpha$
for the exchange contributions are determined by
\begin{equation}
W^{\rm b}_{\lambda} = \lambda \frac{\partial}{\partial\lambda}W^{\rm b} \;\; ,
  \quad
W^{\rm b}_{\lambda\lambda} = \lambda^2 \frac{\partial^2}
               {\partial \lambda^2}  W^{\rm b} \;\; ,  
\end{equation}
\begin{equation}
W^{\rm b}_{\alpha} = \alpha \frac{\partial}{\partial \alpha}  W^{\rm b} \;\; , 
  \quad
W^{\rm b}_{\alpha\alpha} = \alpha^2\frac{\partial^2}{\partial\alpha^2}W^{\rm b}
 \;\; ,  
\end{equation}
\begin{equation}
W^{\rm b}_{\alpha\lambda} = \alpha \lambda \frac{\partial^2}
                       {\partial \alpha \partial \lambda}  W^{\rm b} \;\; .
\end{equation}  
The adiabatic temperature gradient (\ref{Enabla}) expressed by our 
dimensionless potentials (\ref{Efdim1}) and (\ref{Efdim2}) is given by
\begin{equation}  \label{Enabdiml}
\nabla_{\rm ad} = p  k_{\rm T} \frac{\phi_{\rm S}}{c_{\rm P}}
                = \frac{p}{\phi_{\rm S}} ( 1 - \frac{c_{\rm V}}{c_{\rm P}})
\end{equation}
with the isobaric specific heat
\begin{equation}  \label{Ecpdiml}
c_{\rm P} = c_{\rm V} + k_{\rm T} \phi_{\rm S}^{2}
\end{equation}
The Figs.~
\ref{Fnab_H_HE_O_ixr_T79} and ~\ref{Fnab_C_ixr_T5_9} illustrate 
the adiabatic temperature gradient (\ref{Enabdiml}) of the ideal gas 
im comparison with a gas where additionally the exchange term is taken 
into account. Note, that the representation in Figs.~\ref{Fnab_H_HE_O_ixr_T79}
and ~\ref{Fnab_C_ixr_T5_9}  
for extremely high densities serves only to illustrate the asymptotics of the 
theoretical expressions.
Processes relevant to such high densities as, 
e.g., pycnonuclear reactions or electron captures, are not taken into account. 
Fig.~\ref{Fnab_H_HE_O_ixr_T79} illustrates the course of $\nabla_{\rm ad}$ 
for the elements hydrogen, helium and oxygen along a nonrelativistic and a 
weak-relativistic temperature. 
A comparison with the corresponding data for carbon from Lamb (1974) is given 
in Fig.~\ref{Fnab_C_ixr_T5_9}.
Note a disagreement for $T=10^{5}{\rm K}$.
This deviation could be caused by an erroneous constant for the exchange
contribution used by Kovetz et al.\ (1972) and Lamb (1974) and is
discussed in detail in paper I.
A numerical procedure to guarantee a smooth transition between the
approximations given by {\it case a} and {\it case b} is developed in paper I.
The curves as seen in Figs.
\ref{Fnab_H_HE_O_ixr_T79} and ~\ref{Fnab_C_ixr_T5_9}
show how exchange effects vanish with increasing temperature. 
Considering the isotherms in the high density region exchange effects are 
negligible.  
Nevertheless Eqs.~(\ref{Efxb})-(\ref{Ecvxb}) guarantee the accurate asymptotics,
which is a fundamental problem for the calculation of EOS derivatives for
quantities with vanishingly small temperature dependence 
(Miralles \& Van Riper 1996).
\subsection{Correlations}
According to Eq.~(\ref{Eicsredu}) we have to determine the terms labeled 
by the superscript c, which are based on correlation effects. 
As in paper I we aim to present closed-form parametrizations formed
by Pad\'e Approximants to get explicite expressions for the correlation 
contributions of the thermodynamical potentials.
The thermodynamical potentials resulting from the correlation contributions
are characterized by the electronic and ionic Coulomb-coupling 
parameters $\Gamma_{\rm e}=(4 \pi n_{\rm e}/3)^{1/3} e^{2}/kT$ 
and $\Gamma_{\rm i}=\Gamma_{\rm e}\langle Z^{5/3} \rangle$ giving
the strength of the Coulomb interaction, and by the degeneracy parameter 
$\Theta=T/T_{\rm F}=2(4 /9 \pi)^{2/3} r_{\rm s}/\Gamma_{\rm e}$ 
describing the quantum state of the system.
The parameter 
$r_{\rm s}=m_{\rm e} e^{2}/\hbar^{2} (3/4 \pi n_{\rm e})^{1/3}$ is the ratio
of the mean interelectronic distance to the (electronic) Bohr-radius.  
\subsubsection{Electron-electron interaction} \label{eesec}
The Pad\'e Approximants for free Helmholtz- and Gibbs energies of the 
nonrelativistic electronic subsystem at arbitrary degeneration and Coulomb 
coupling have the general structure   
\begin{equation}  \label{Efeec}
f_{\rm ee}^{\rm c} =
   - \frac{ a_{0} \Gamma_{\rm e}^{3/2} -  a_{2} \Gamma_{\rm e}^{6}
              \left[ \varepsilon_{\rm c}(r_{\rm s}, 0) 
   +          \Delta \varepsilon_{\rm c}(r_{\rm s}, \tau) \right]  / \tau }
       {1 + a_{1} \Gamma_{\rm e}^{3/2} +  a_{2} \Gamma_{\rm e}^{6}} \;\; ,
\end{equation}
\begin{equation}  
g_{\rm ee}^{\rm c} =
   - \frac{ s_{0} \Gamma_{\rm e}^{3/2} -  s_{2} \Gamma_{\rm e}^{6}
              \left[ \mu_{\rm c}(r_{\rm s}, 0)
   +          \Delta \mu_{\rm c}(r_{\rm s}, \tau) \right] / \tau }
       {1 + s_{1} \Gamma_{\rm e}^{3/2} +  s_{2} \Gamma_{\rm e}^{6}}  \;\; ,
\end{equation}
with ground-state energies 
$\varepsilon_{\rm c}(r_{\rm s}, 0)$ and $\mu_{\rm c}(r_{\rm s}, 0)$ 
derived by Vosko \etal\ (1980) and low-temperature corrections 
$\Delta \varepsilon_{\rm c}(r_{\rm s}, \tau)$ and 
$\Delta \mu_{\rm c}(r_{\rm s}, \tau)$, which are given in paper I. 
The coefficients read
\begin{equation}
a_{0} = \frac{1}{\sqrt{3}} f_{0}(\Gamma_{\rm e}) \;,
\quad\quad\quad\quad\quad\quad
s_{0} = \frac{3}{2} a_{0}
      + \frac{1}{3} \Gamma_{\rm e} \frac{{\rm d} a_{0}}{{\rm d} \Gamma_{\rm e}}
\end{equation}
\begin{equation}
f_{0}(\Gamma) = \frac{1}{2} \left( \frac{1}
                            {(1 + 0.1088 \Gamma_{\rm e})^{1/2}} +
                \frac{1}{(1 + 0.3566 \Gamma_{\rm e})^{3/2}} \right)
\end{equation}
\begin{equation}    \label{Ea1KSTAR}
a_{1} = \frac{3 \sqrt{3}}{32} \sqrt{2 \pi \tau}
                \left[1+K_{\rm e}^{\ast}\left(\sqrt{2/\tau}\right)\right] \; , 
\quad\quad 
s_{1} = \frac{8}{9} a_{1}
\end{equation}
\begin{equation}
a_{2} = 6 \frac {\tau}{r_{\rm s}^{2}}
\; , \quad\quad\quad\quad\quad\quad\quad\quad\quad\quad\quad\quad \;\;\;
s_{2} = a_{2} \;\;\;  .
\end{equation}  
Note that we have modified the coefficients $a_{1}$ and $s_{1}$ as compared
to paper I
by a rearrangement of the quantum virial function (Ebeling \etal\ 1976)
\begin{equation} \label{Equavirf}
K_{\rm e}^{\ast}(x)  =  E_{2}^{\ast}(-x)
  - \frac{8x}{\sqrt{\pi}} \left[ Q_{3}^{\ast}(-x)
  - \frac{1}{6} \ln |x| - C_{0} \right]
\end{equation}
In paper I we incorporated for $K_{\rm e}^{\ast}(x)$ in
Eq.~(\ref{Ea1KSTAR}) only the first part of Eq.~(\ref{Equavirf}) as  
proposed by Ebeling (1993). 
The second term was considered by the coefficient $c_{1}$ of the ion-electron 
contribution (see Eq.~(85) in paper I).
The justification to rearrange these summations over the electrons is
the omission of the coefficient $a_{3}$ in Eq.~(53) in paper I, 
which has only been introduced to optimize the interpolation.
For $E_{2}^{\ast}$ and $Q_{3}^{\ast}$ we derived Pad\'e approximations
as given in paper I with improvements to achieve higher accuracy.
\begin{equation}
E_{2}^{\ast} (-x) =  \frac{ \ln 2 + 0.0113 x^{2} + 0.1 x^{5}
         \left( \frac{2}{\sqrt{\pi} x} -
         \frac{1}{x^{2}} \right) } { 1 + \pi^{3/2}x/(18 \ln 2)  + 0.1 x^{5} }
\end{equation}
\begin{equation}
Q_{3}^{\ast} (-x) =   \frac{ C_{1} + x^{5}
\left(\frac{1}{6}\ln |x|+C_{0}+\frac{\sqrt{\pi}}{8x}-\frac{1}{9 x^{2}}\right)}
 { 1 - \frac{9}{10 C_{1}} \left( \frac{\pi^{5/2} x}{192}
                    - \frac{ x^{2}}{75} \right)+x^{5}} \;\; 
\end{equation}
with $C_{0}=\frac{1}{6}\left(\ln 3+    2 C_{\rm E}    -\frac{11}{6}\right)$
and  $C_{1}=\frac{1}{6}\left(\ln 3+\frac{C_{\rm E}}{2}-\frac{ 1}{2}\right)$.

A detailed comparison of the Pad\'e formula (\ref{Efeec}) with
most advanced many body calculations is given in paper I and recently
in Stolzmann \& R\"osler (1998b).
\begin{figure}
\centering
\epsfxsize=0.49\textwidth
\mbox{\epsffile{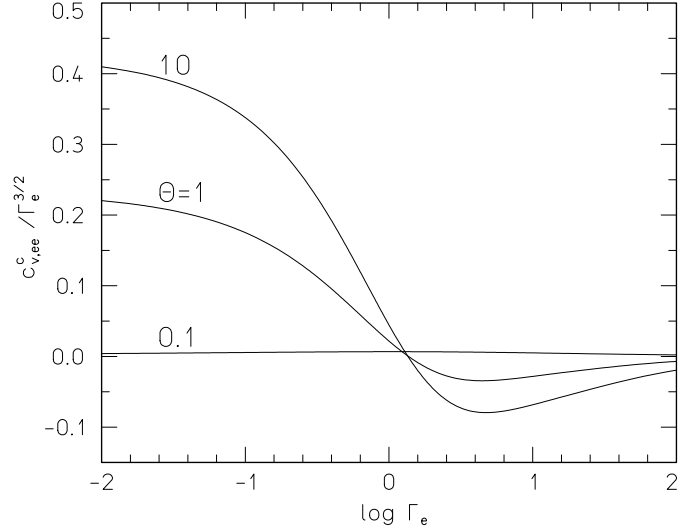}}
\caption[cv_ee_3thel]
{The isochoric specific heat for the electronic subsystem
 at different degeneracy
 ($\Theta=10$ (classical), $1$ (intermediate), $0.1$ (strong))
 vs.\ Coulomb-coupling parameter.}
                              \label{Fcv_ee_3thel}
\end{figure}
The electronic correlation contribution for the pressure is obtained by 
\begin{equation}  \label{Epeec}
p_{\rm ee}^{\rm c} = g_{\rm ee}^{\rm c} - f_{\rm ee}^{\rm c}  \;\; ,
\end{equation}
The internal energy can be described by the Pad\'e Approximant
\begin{equation}
u_{\rm ee}^{\rm c} =
    - \frac{ u_{0} \Gamma_{\rm e}^{3/2} -  u_{2} \Gamma_{\rm e}^{6}
      \left[ \varepsilon_{\rm c}(r_{\rm s}, 0)
    + \Delta \varepsilon_{\rm c}(r_{\rm s}, \tau) \right] / \tau } 
        {1 + u_{1} \Gamma_{\rm e}^{3/2} +  u_{2} \Gamma_{\rm e}^{6}} \;\; ,
\end{equation}
with the coefficients
\begin{equation}
u_{0} = \frac{3}{2} a_{0}
      + \Gamma_{\rm e} \frac{{\rm d} a_{0}}{{\rm d} \Gamma_{\rm e}} \; ,
 \quad\quad\quad\quad\quad\quad
u_{2} = a_{2}
\end{equation}
\begin{equation}
u_{1} = 2 \left(a_{1}-\frac{\tau}{3}\frac{{\rm d}a_{1}}{{\rm d}\tau}\right)   
      =   \frac{5}{3}a_{1}-\frac{\sqrt{3}}{16} \sqrt{2 \pi \tau}
      \left(\tau  \frac{{\rm d}}{{\rm d}\tau} K_{\rm e}^{\ast}(x)\right)
\end{equation}  
The temperature derivative of the quantum virial function (\ref{Equavirf})
is given by
\begin{eqnarray}
\tau \frac{{\rm d}}{{\rm d}\tau} K_{\rm e}^{\ast}(x) & = &
           \tau \frac{{\rm d}}{{\rm d}\tau} E_{2}^{\ast}(-\tau)
  - \frac{1}{2} K_{\rm e}^{\ast}(-\tau) + \frac{1}{2} E_{2}^{\ast}(-\tau)
             \nonumber \\
 & &
  -  \frac{16}{\sqrt{2 \pi \tau}}
\left[ \tau \frac{{\rm d}}{{\rm d}\tau}Q_{3}^{\ast}(-\tau)+\frac{1}{12} \right]
 \;\; .
\end{eqnarray}
For the temperature derivatives of $E_{2}^{\ast}$ and $Q_{3}^{\ast}$ we
derive the Pad\'e Approximants
\begin{equation}
           \tau \frac{{\rm d}}{{\rm d}\tau} E_{2}^{\ast}(-\tau)  =
    \frac{ \frac{\pi^{3/2}}{36} x +  \frac{x^{5}}{12}
    \left( \frac{1}{\sqrt{\pi} x} - \frac{1}{x^{2}} \right)}
 {1 + 0.113 \frac{36}{\pi^{3/2}} x \left(1 - 0.473 x \right)+\frac{x^{5}}{12}}
\end{equation}
\begin{equation}  \label{Ederiq3ast}
           \tau \frac{{\rm d}}{{\rm d}\tau} Q_{3}^{\ast}(-\tau)  =
      \frac{- \frac{\pi^{5/2}}{384} x + 0.1 x^{4.5} \left(
   -  \frac{1}{12} + \frac{\pi^{1/2}}{16x} - \frac{1}{6 x^{2}} \right)}
     { 1 + \frac{384}{25 \pi^{5/2}} x + 0.1 x^{4.5}} \; .
\end{equation}
The correlation contribution $1/k_{\rm T,ee}^{\rm c}$ for the bulk modulus 
is given by
\begin{equation}
\frac{1}{k_{\rm T,ee}^{\rm c}} = \frac{ k_{0} \Gamma_{\rm e}^{3/2} 
                                      + k_{2} \Gamma_{\rm e}^{6}
           \left[ p_{\rm c}(r_{\rm s}, 0) 
   \left( 1 - 1 / k_{\rm c}(r_{\rm s}, 0) \right) \right] / \tau }
                                   {1 + k_{1} \Gamma_{\rm e}^{3/2} 
                                      + k_{2} \Gamma_{\rm e}^{6}}
\end{equation}
with the coefficients
\begin{equation}
k_{0} = -\frac{3}{4} a_{0}  - \frac{\Gamma_{\rm e}}{4}
       \frac{{\rm d} a_{0}}{{\rm d} \Gamma_{\rm e}}  \; ,
                      \quad\quad
k_{2} = 10 a_{2} \; , \quad\quad
k_{1} = 30 a_{1} \; .
\end{equation}
%
The ground-state bulk modulus is given by
\begin{equation}
\frac{1}{k_{\rm c}(r_{\rm s}, 0)} = \frac{3}{0.062181} \; \frac{r_{\rm s}}{3}
                \frac{{\rm d}}{{\rm d}r_{\rm s}} p_{\rm c}(r_{\rm s}, 0) \;\; .
\end{equation}
Using the Monte-Carlo data-fit from  Vosko \etal\ (1980) for 
$p_{\rm c}(r_{\rm s}, 0)$ (see paper I) 
we apply 
\begin{equation}
\frac{1}{k_{\rm c}(r_{\rm s},0)}  =  
                     \left[ \frac{\frac{1}{2} m_{1} \sqrt{r_{\rm s}}}
                                         {1 + m_{1} \sqrt{r_{\rm s}}}
                          - \frac{\frac{1}{2} m_{1} \sqrt{r_{\rm s}}
                    +  m_{2} r_{\rm s} + \frac{3}{2}m_{3} r_{\rm s}^{3/2}}
      {1 + m_{1} \sqrt{r_{\rm s}}+m_{2} r_{\rm s} + m_{3} r_{\rm s}^{3/2}}
                     \right]
\end{equation}  
The coefficient of strain $\phi_{\rm S,ee}^{\rm c}$ can be expressed by
\begin{eqnarray}
\phi_{\rm S,ee}^{\rm c} & = & p_{\rm ee}^{\rm c} + u_{\rm ee}^{\rm c}
  \nonumber \\
 & & \;\;
 +  \frac{ \phi_{0} \Gamma_{\rm e}^{3/2} -  \phi_{2} \Gamma_{\rm e}^{6}
           \left[ \mu_{\rm c}(r_{\rm s}, 0) - \frac{3}{2}
             \Delta \varepsilon_{\rm c}(r_{\rm s}, \tau)  \right] / \tau }
      {1 + \phi_{1} \Gamma_{\rm e}^{3/2} +  \phi_{2} \Gamma_{\rm e}^{6}}
\end{eqnarray}
with the coefficients
\begin{equation}
\phi_{0} = \frac{9}{4} a_{0} + \frac{3}{4} \Gamma_{\rm e}
   \frac{{\rm d} a_{0}}{{\rm d} \Gamma_{\rm e}} \; ,
 \quad\quad\quad\quad
\phi_{2} = \frac{1}{2}  a_{2} \; ,
\end{equation}
\begin{equation}
\phi_{1} = \frac{16}{9} a_{1} - \frac{2}{3} \tau \frac{{\rm d} a_{1}}
                                                        {{\rm d} \tau}
\end{equation}
and for the isochoric specific heat $c_{\rm V,ee}^{\rm c}$
the Pad\'e Approximant can be obtained by
\begin{equation}  \label{Ecveecpad}
c_{\rm V,ee}^{\rm c} =
  \frac{ v_{0} \Gamma_{\rm e}^{3/2} -  v_{2} \Gamma_{\rm e}^{6}
   \left[ 2 f(r_{\rm s}) \Theta^{2} \left(1 + \frac{2}{9} \ln \Theta \right)
  \right] / \tau }{1 + v_{1} \Gamma_{\rm e}^{3/2} +  v_{2} \Gamma_{\rm e}^{6}}
\end{equation}
with the coefficients
\begin{equation}
v_{0} = \frac{3}{4} \left[ a_{0} + \frac{1}{4} \Gamma_{\rm e}
        \frac{{\rm d} a_{0}}{{\rm d} \Gamma_{\rm e}}
                   \right]  \; ,  \quad\quad\quad\quad\quad
v_{2} =  a_{2} \;,
\end{equation}
\begin{eqnarray}
v_{1} & = &  \frac{3 \sqrt{3}}{32} \sqrt{2 \pi \tau} \;
 \left[ 1 + \frac{7}{12}
 \left(1 - \frac{\sqrt{2 \pi \tau}}{16}K_{\rm e}^{\ast}(-\tau)\right)
       \right.  \nonumber \\
  & &  \left.   
+ \frac{1}{24} \left( 1 + E_{2}^{\ast}(-\tau) \right)
- \frac{1}{2}  \left(\tau \frac{{\rm d}}{{\rm d}\tau} E_{2}^{\ast}(-\tau)
               \right)
       \right.  \nonumber \\
  & &  \left. 
- \frac{5}{6} \frac{16}{\sqrt{2 \pi \tau}}
\left(\tau \frac{{\rm d}}{{\rm d}\tau}Q_{3}^{\ast}(-\tau)+\frac{1}{12}\right)
        \right] \;\; .
\end{eqnarray}
Fig.~\ref{Fcv_ee_3thel} shows that the electronic correlation part of the 
isochoric specific heat can be neglected in the case of strongly degenerate
electrons.
\subsubsection{Ion-ion interaction}
The classical one-component plasma (OCP) is a thoroughly studied plasma 
system and the results are unified in closed-form parametrizations 
(Hansen 1973, Graboske \etal\ 1975, Brami \etal\ 1979, Ebeling \& Richert 1985, 
 Ichimaru 1993, Kahlbaum 1996, Chabrier \& Potekhin 1998). 
For the ion-ion correlation we modify the Pad\'e Approximants
used in paper I by the new formula (Stolzmann \& Bl\"ocker 1998, 1999) for 
the free energy
\begin{equation}  \label{Efiic}
 f_{\rm ii}^{\rm c}= 
  - \frac{b_{0} \Gamma_{\rm i}^{3/2} \left[ 1 + 
          b_{3} \Gamma_{\rm i}^{3/2} F(\Gamma_{\rm i}) \right]
       +  b_{2} \Gamma_{\rm i}^{6}\varepsilon_{\rm ii}(\Gamma_{\rm i})} 
     {1 - b_{1} \Gamma_{\rm i}^{3}   G(\Gamma_{\rm i})
       +  b_{2} \Gamma_{\rm i}^{6}} \; ,
\end{equation}
with $F(\Gamma_{\rm i}) = \ln \Gamma_{\rm i} + B_{0}$ and 
$G(\Gamma_{\rm i}) = \ln \Gamma_{\rm i} + B_{1}$
\begin{equation}
B_{0} = \frac{2}{3}\left(2C_{\rm E}+\frac{3}{2}\ln 3-\frac{11}{6}\right) \; ,
\end{equation}
\begin{equation}
B_{1} = \frac{2}{3}\left(2C_{\rm E}+\frac{1}{2}\ln 3+2\ln 2-\frac{17}{6}\right)
        -0.4765
\end{equation}
and the coefficients
\begin{equation}
b_{0} = \frac{1}{\sqrt{3}} \frac{ \langle Z^{2 }  \rangle^{3/2}}
         {\langle Z \rangle^{1/2} \langle Z^{5/3} \rangle^{3/2}}
\; , 
\quad\quad\quad  
b_{1} = \frac{3 \sqrt{3}}{8b_{0}} \; ,
\end{equation}
\begin{equation}
b_{2} = 100  \; ,
\quad\quad\quad
b_{3} = \frac {b_{1}}{\sqrt{3}}  \; ,
\end{equation}
Eq.~(\ref{Efiic}) is based on the classical $\Gamma_{\rm i} < 1$ result from
Cohen \& Murphy (1969) and for $\Gamma_{\rm i} \ga 1$ we take into
account the most recent Monte-Carlo fit for the free energy
$\varepsilon_{\rm ii} (\Gamma_{\rm i})$ of the liquid
OCP from DeWitt \& Slattery (1999)
\begin{eqnarray}
\varepsilon_{\rm ii}(\Gamma_{\rm i}) & = &
                          0.899172 \Gamma_{\rm i}
                       +  0.274823 \ln  \Gamma_{\rm i}
 \nonumber \\
  & &  \quad\quad\quad\quad\quad
                       -  1.864179   \Gamma_{\rm i}^{\; 0.323064}
                       +  1.4018    \;\; .
\end{eqnarray}
We remark, that in our notation of Eq.~(\ref{Eicsredu}) for 
$\varepsilon_{\rm ii}(\Gamma_{\rm i})$ we have to take for 
$\Gamma_{\rm i} \ga 178$
\begin{eqnarray}
\varepsilon_{\rm ii}(\Gamma_{\rm i}) & = &
                          0.895929 \Gamma_{\rm i}
                       -  1.5 \ln  \Gamma_{\rm i}
                       +  3.9437   \Gamma_{\rm i}^{-1}
 \nonumber \\
  & &  \quad\quad\quad\quad\quad
                       +  \; 1245     \Gamma_{\rm i}^{-2}
                       +  1.1703
\end{eqnarray}
which, with  Eq.~(\ref{Eionid}) rewritten by  
\begin{equation}  \label{Eionid2}
f_{\rm i}^{\rm id} =   \frac{3}{2} \ln \frac{kT}{\rm {Ryd_{i}}} 
                     + 3 \ln \Gamma_{\rm i} - 0.7155  ,
\end{equation}
delivers the thermal (solid phase) energy of the OCP given by
Hansen (1972), Pollock \& Hansen (1973), Slattery \etal\ (1980), and 
Stringfellow \etal\ (1990). The quantity
${\rm Ryd_{\rm i}}=A Z^{4} m_{\rm p} / m_{\rm e} {\rm Ryd}$ denotes the 
ionic Rydberg energy with the (electronic) Rydberg unit 
${\rm Ryd} =m_{\rm e} e^4 / 2 \hbar^{2}$.

The free Gibbs energy $g_{\rm ii}^{\rm c}$ and the isochoric specific heat
$c_{\rm V,ii}^{\rm c}$ are expressed by the Pad\'e Approximants
\begin{equation}  \label{Egiic}
 g_{\rm ii}^{\rm c} =
-\frac{\frac{3}{2} b_{0} \Gamma_{\rm i}^{3/2} \left[ 1 +
       \frac{4}{3} b_{3} \Gamma_{\rm i}^{3/2}
                \left( F(\Gamma_{\rm i})  + \frac{1}{6} \right) \right]
                +  b_{2} \Gamma_{\rm i}^{6} \mu_{\rm ii}(\Gamma_{\rm i})}
  {1 - \frac{5}{3} b_{1} \Gamma_{\rm i}^{3}
                 \left(G(\Gamma_{\rm i})  + \frac{2}{15} \right )
  \left[ 1 + \frac{3}{4} \Gamma_{\rm i}^{3/2} \right]
                +  b_{2} \Gamma_{\rm i}^{6}}
\end{equation}
\begin{equation}     \label{Ecviicpad}
c_{\rm V,ii}^{\rm c} =
 \frac{\frac{3}{4} b_{0} \Gamma_{\rm i}^{3/2} \left[ 1 +
                 8 b_{3} \Gamma_{\rm i}^{3/2}
                 \left(F(\Gamma_{\rm i})  + \frac{5 }{6 } \right ) \right]
               + 6 b_{2} \Gamma_{\rm i}^{6} \varrho_{\rm ii}(\Gamma_{\rm i})}
           { 1 - 7 b_{1} \Gamma_{\rm i}^{3}
                 \left(G(\Gamma_{\rm i})  + \frac{32}{63} \right )
               + 6 b_{2} \Gamma_{\rm i}^{6}}
\end{equation}
with
\begin{equation}
\mu_{\rm ii}(\Gamma_{\rm i}) = \varepsilon_{\rm ii}(\Gamma_{\rm i}) +
                               \frac{1}{3} \Gamma_{\rm i}
 \frac{\rm d \varepsilon_{\rm ii} (\Gamma_{\rm i})}{\rm d \Gamma_{\rm i}}
\; , \quad
\varrho_{\rm ii}(\Gamma_{\rm i}) = \Gamma_{\rm i}^{2}
    \frac{{\rm d}^{2} \varepsilon_{\rm ii}(\Gamma_{\rm i})}
                                  {{\rm d} \Gamma_{\rm i}^{2}} \; .
\end{equation}
The pressure, internal energy, isothermal compressibility, 
and coefficient of strain (Hansen 1973, Stolzmann \& Bl\"ocker 1999) 
are calculated by 
\begin{equation}   \label{Epiicpad}
p_{\rm ii}^{\rm c} = g_{\rm ii}^{\rm c}-f_{\rm ii}^{\rm c}   \;\;,
\end{equation}
\begin{equation}   \label{Euiicpad}
u_{\rm ii}^{\rm c}  = 3 \; p_{\rm ii}^{\rm c}  \;\;,
\end{equation}
\begin{equation}  \label{Ecomiiocp}
\frac{1}{k_{\rm T,ii}^{\rm c}} = - \frac{1}{9} \; c_{\rm V,ii}^{\rm c}
                                 + \frac{4}{9} \; u_{\rm   ii}^{\rm c} \;\;,
\end{equation}
\begin{equation}  \label{Eptiiocp}
\phi_{\rm S,ii}^{\rm c} = \frac{1}{3} \; c_{\rm V,ii}^{\rm c}  \;\;.
\end{equation}  
\begin{figure}
\centering
\epsfxsize=0.49\textwidth
\mbox{\epsffile{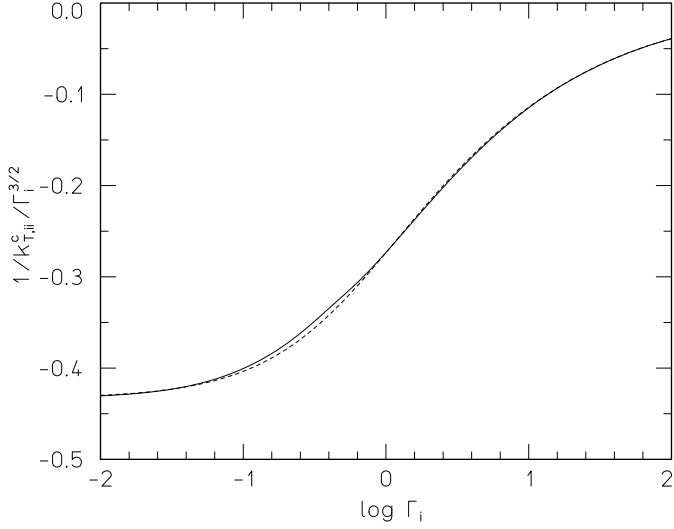}}
\caption[k_ii_gam]
{The compressibility for the ionic subsystem vs.\ coupling parameter
 $\Gamma_{\rm i}$.
 The solid line represents the correlation contribution (\ref{Ecomiiocp})
 and the dashed line refers to the parametrization from Hansen (1973).}
                                         \label{Fk_ii_gam}
\end{figure}
Fig.~\ref{Fk_ii_gam} compares our Pad\'e Approximants of the isothermal 
compressibility with the fit formulae from Hansen (1973).
Further comparisons between different interpolation formulae for the 
classically ionic subsystem, e.g. from Hansen (1973), Ichimaru (1994), 
Kahlbaum (1996), Chabrier \& Potekhin (1998) 
can be found in Stolzmann \& Bl\"ocker (1999).
 
Ionic quantum effects, which were included approximately in our earlier Pad\'e 
Approximants for the ionic subsystem (see paper I) will be considered 
separately in the following section.  
\subsubsection{Ionic quantum effects}
The classical description for the subsystem of ionic OCP becomes inaccurate 
in the region of high densities at moderate temperatures.
A measure for the quantum effects of the ions is the parameter 
(weight fraction $X_{i}$, molecular weight $A_{i}$, proton mass $m_{\rm p}$,
 atomic mass unit $m_{\rm u}=1.66053 \cdot 10^{-24}{\rm g}$)
\begin{equation}  \label{Eionqupara}
\Theta_{\rm i} = \frac{\hbar \omega_{\rm p}}{kT} =
 \frac{4 \sqrt{\pi}}{kT} \left( \frac{\rho}{m_{\rm u}} 
 \frac{m_{\rm e}}{m_{\rm p}}
        \left( \frac{\hbar^{2}}{m_{\rm e} e^{2}} \right)^{3}
        \sum_{i} \frac{Z_{i}^{2} X_{i}}{A_{i}^{2}} \right)^{1/2} 
\end{equation}
which is related by
\begin{equation}  \label{Eionquparag}
\Theta_{\rm i} = \sqrt{\frac{3}{R_{\rm S}}} \Gamma_{\rm i} \;\;,
\end{equation}
and whereas $\omega_{\rm p}$ denotes the (ionic) plasma frequency and 
$R_{\rm S}=m_{\rm i} Z^{2}e^{2}/\hbar^{2} (3/4 \pi n_{\rm i})^{1/3}$ 
is the ratio of the (ionic) Wigner-Seitz-radius to the (ionic) Bohr-radius.
\begin{figure}
\centering
\epsfxsize=0.49\textwidth
\mbox{\epsffile{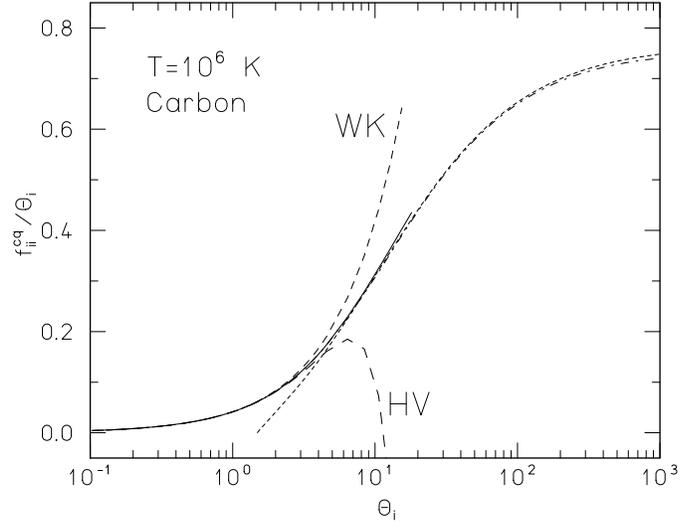}}
\caption[f_qc_ii_T6]
{The free energy contribution of the ionic quantum corrections 
 for carbon at $T=10^{6} {\rm K}$ vs.\ ionic quantum parameter 
 (Eq.~\ref{Eionqupara}).
 The solid line represents Nagara \etal\ (1987) (Eq.~\ref{Efiicq}) 
 (shown only within its validity regime), 
 the dashed line refers to Chabrier \etal\ (1992), and the
 dashed-dotted line to Iyetomi \etal\ (1993). The curve labeled by
 WK refers to the first order in the Wigner-Kirkwood expansion and 
 HV to Hansen \& Vieillefosse (1975).}                
                            \label{Ff_qc_ii_T6}
\end{figure}
\begin{figure}
\centering
\epsfxsize=0.49\textwidth
\mbox{\epsffile{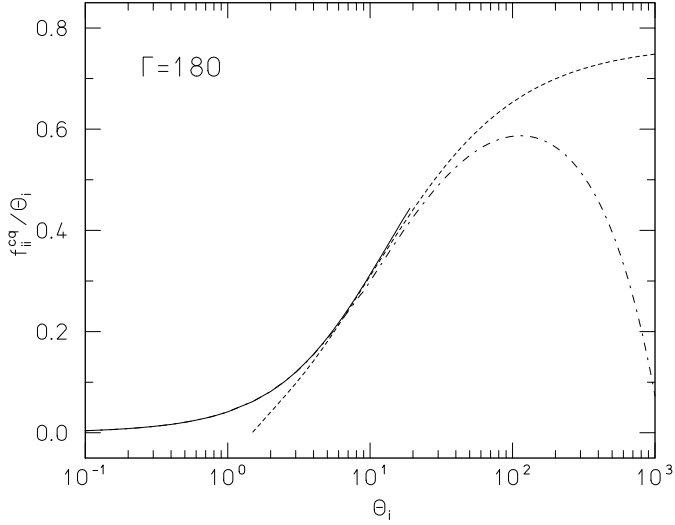}}
\caption[f_qc_ii_G180]
{The free energy contribution of the ionic quantum corrections 
 at $\Gamma_{\rm i}=180$ vs.\ $\Theta_{\rm i}$. 
 The solid line represents Nagara \etal\ (1987) (Eq.~\ref{Efiicq})
 (shown only within its validity regime), 
 the dashed line refers to Chabrier \etal\ (1992), and the
 dashed-dotted line to Iyetomi \etal\ (1993).}      
                                         \label{Ff_qc_ii_G180}
\end{figure}  
Quantum corrections for a solid OCP over a broad region of $\Theta_{\rm i}$ 
were calculated first by Pollock \& Hansen (1973).  
Chabrier \etal\ (1992) (see also Chabrier 1993) and Iyetomi \etal\ (1993)
derived an free energy for a solid OCP model with quantum corrections.   
Nagara \etal\ (1987) calculated quantum corrections for the free energy
of the OCP valid for fluid and solid phases by a new expansion method.
In our model ionic quantum effects according to Nagara \etal\ (1987) 
as well as Chabrier \etal\ (1992) will be included.
 
Following paper I, the free energy calculated within the 
so-called ``sixth reduced moment approximation'' of Nagara \etal\ (1987) 
is given by
\begin{equation}    \label{Efiicq}
f_{\rm ii}^{\rm cq}    =  3 \left[ q_{0}(t) + z_{4} q_{4}(t)  
       - \frac{3}{4} \frac{t}{\Gamma_{\rm i}} z_{5} q_{5}(t)
                                            + z_{6} q_{6}(t) \right]  
\end{equation}
with $t=\Theta_{\rm i} / 2 \sqrt{3}$. 
Generally, the reduced moments $z_{i}$ are dependent of $\Gamma_{\rm i}$. 
For simplicity we choose for the solid phase the values 
for the rigid bcc lattice (Nagara \etal\ 1987):
\begin{equation}
z_{4} = 0.827702 \;\; , \;\;\;\; 
z_{5} = 1.131    \;\; , \;\;\;\; 
z_{6} = 0.55045 \;\; .
\end{equation}
The functions $q_{i}(t)$ are given by 
\begin{equation}
q_{0}(t) = \ln \left( \frac{\sinh t}{t} \right) 
\end{equation}
\begin{equation}
q_{4}(t) =   \frac{1}{4}  -  \frac{t}{8} \coth t 
           - \frac{t^{2}}{8} \frac{1}{\sinh^{2} t}
\end{equation}
\begin{equation}
q_{5}(t) =   \frac{t}{3} \coth^{2} t - \frac{3}{2} \coth t 
           - \frac{5}{6} t \frac{1}{\sinh^{2} t} + \frac{2}{t} 
\end{equation}
\begin{equation}
q_{6}(t) = - \frac{1}{6} + \frac{t}{16} \coth t 
           + \frac{t^{2}}{16} \frac{1}{\sinh^{2} t} 
             \left( 1 + \frac{2 t}{3} \coth t \right)  \;\; .
\end{equation}
\begin{figure}
\centering
\epsfxsize=0.49\textwidth
\mbox{\epsffile{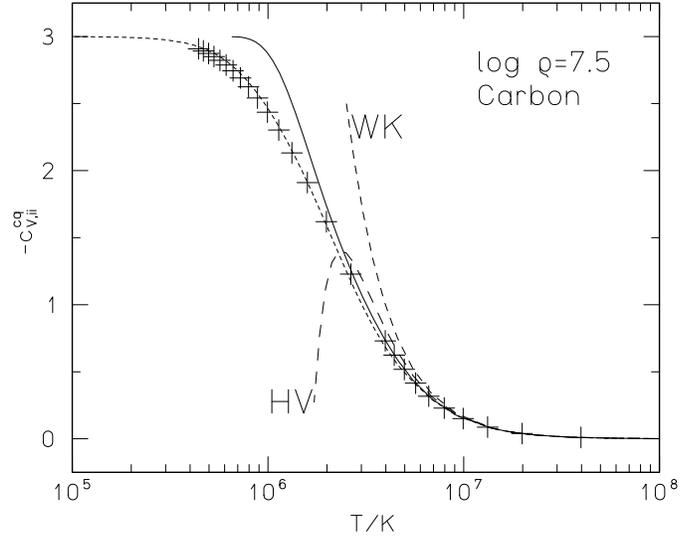}}  
\caption[cv_cqii_r75_C]
{The isochoric specific heat for the ionic quantum corrections 
 for carbon at $\rho=10^{7.5} \rm {g/cm^{3}}$ vs.\ temperature.
 The solid line represents Nagara \etal\ (1987) (Eq.~\ref{CViicq})
 (shown only within its validity regime), 
 the dashed line refers to Chabrier (1993), and
 the crosses to  Pollock \& Hansen (1973). 
 The curve labeled WK refers to the first order in the 
 Wigner-Kirkwood expansion and HV to Hansen \& Vieillefosse (1975).}     
                                       \label{Fcv_cqii_r75_C}
\end{figure}
\begin{figure}
\centering
\epsfxsize=0.49\textwidth
\mbox{\epsffile{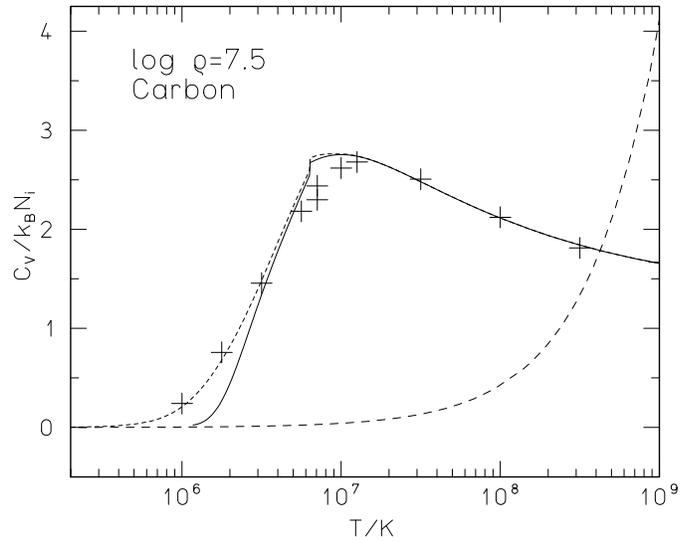}}
\caption[cv_r75_C_nnn_cad]
{The electronic and ionic parts of the isochoric specific heat 
 for carbon at $\rho=10^{7.5} \rm {g/cm^{3}}$ vs.\ temperature.
 The long-dashed line represents the electronic terms of Eq.~(\ref{Eicsredu}) 
 The solid line (applies Nagara \etal\ (1987) for $c_{\rm V,ii}^{\rm cq}$) 
 and the short-dashed line (applies Chabrier (1993) for 
 $c_{\rm V,ii}^{\rm cq}$) represent the ionic terms of Eq.~(\ref{Eicsredu}).
 The crosses refer to the ionic contribution calculated by Lamb \& Van Horn 
 (1975).}                             \label{Fcv_r75_C_nnn_cad}
\end{figure}
The free energy expansion (\ref{Efiicq}) reproduces the Wigner-Kirkwood 
expansion (Wigner 1932, Kirkwood 1933) up to the order $\Theta_{\rm i}^{6}$.
Furthermore the expansion of Nagara \etal\ (1987) delivers 
the high-$\Theta_{\rm i}$ asymptotics of the Coulomb lattice by 
\begin{equation}
f_{\rm ii}^{\rm cq} = \frac{3}{2} \mu_{1} \Theta_{\rm i}
\end{equation}
with $\mu_{1} = 0.564$. 
For the harmonic lattice approximation one obtains $\mu_{1} = 0.511$ 
(Hansen \& Vieillefosse 1975). 
\\
The ionic quantum effects for our set of potentials are given by the 
resulting formulae 
\begin{eqnarray}
p_{\rm ii}^{\rm cq} & = & \frac{3}{2} t 
  \left[ q_{0}^{\prime}(t)+z_{4}q_{4}^{\prime}(t)+z_{6}q_{6}^{\prime}(t) 
              -\frac{3}{4}\frac{t}{\Gamma_{\rm i}}z_{5}q_{5}^{\prime}(t)
  \right]  \nonumber  \\    
   & & \quad\quad\quad\quad\quad\quad\quad\quad\quad\quad\quad \;
              -\frac{3}{8}\frac{t}{\Gamma_{\rm i}} z_{5}q_{5}(t) 
\end{eqnarray}
\begin{equation}
g_{\rm ii}^{\rm cq}  =  p_{\rm ii}^{\rm cq} + f_{\rm ii}^{\rm cq} 
\end{equation} 
\begin{equation}
u_{\rm ii}^{\rm cq} = 2 p_{\rm ii}^{\rm cq} 
            + \frac{3}{4}\frac{t}{\Gamma_{\rm i}} z_{5} q_{5}(t)
\end{equation}
\begin{equation}   \label{CViicq}
c_{\rm V,ii}^{\rm cq} = - 3 t^{2} \left[ q_{0}^{\prime\prime}(t) 
                                 + z_{4} q_{4}^{\prime\prime}(t)  
            - \frac{3}{4}\frac{t}{\Gamma_{\rm i}} 
                                   z_{5} q_{5}^{\prime\prime}(t)
                                 + z_{6} q_{6}^{\prime\prime}(t) 
                                  \right]
\end{equation}
\begin{equation}
\frac{1}{k_{\rm T,ii}^{\rm cq}} = \frac{1}{4} \left[ 3 u_{\rm ii}^{\rm cq} 
                         - c_{\rm V,ii}^{\rm cq}
                         - \frac{3}{4}\frac{t}{\Gamma_{\rm i}} z_{5}
                           \left(\frac{7}{3} q_{5}(t) + 2 t q_{5}^{\prime}(t) 
                           \right) \right]
\end{equation}
\begin{equation}
\phi_{\rm S,ii}^{\rm cq} = \frac{1}{2} \left[ c_{\rm V,ii}^{\rm cq}
                         - \frac{3}{4}\frac{t}{\Gamma_{\rm i}} z_{5}     
                         \left(\rule{0mm}{5mm} q_{5}(t) - t q_{5}^{\prime}(t)
                           \right) \right]                          
\end{equation}
The functions $q_{i}^{\prime}(t)$ and $q_{i}^{\prime\prime}(t)$ are first and 
second derivatives of the functions $q_{i}(t)$ with respect to $t$.

Although the expansion of Nagara \etal\ (1987) can be applied to 
estimate quantum effects for fluid ionic phases as well as for solid phases
this expansion is restricted to $\Theta_{\rm i} \la 17$, which is 
equivalent to $\Gamma_{\rm i} \la 4 \cdot 10^{3} (A Z^{4}/(T/{\rm K}))^{1/3}$
or to $\rho \la 5 \cdot 10^{-6} (T/{\rm K}))^{2} (A / Z)^{2}$ for the 
density-temperature regime.
For $\Gamma_{\rm i} \approx \Gamma_{\rm m} \ga 180$ it is convenient to use 
the quantum crystal models from Iyetomi \etal\ (1993) or from 
Chabrier \etal\ (1992), which are applicable for a wide region of 
$\Theta_{\rm i}$. 
\\
If we extract the ionic quantum corrections for the free energy of a crystal 
from Chabrier \etal\ (1992) one obtains (Stolzmann \& Bl\"ocker 1994)
\begin{eqnarray}    \label{Efiicqcad}
f_{\rm ii}^{\rm cq} & = & - \frac{2}{3}  D_{\rm 3}(\alpha \Theta_{\rm i})
                +2\ln\left[1-{\rm e}^{-\alpha \Theta_{\rm i}}\right]
                + \ln\left[1-{\rm e}^{-\gamma \Theta_{\rm i}}\right]
 \nonumber \\
 & & \quad\quad\quad\quad\quad\quad
    + \beta \Theta_{\rm i} - 3 \ln \left(\delta\Theta_{\rm i}\right) 
\end{eqnarray}
with (Chabrier 1993) 
\begin{equation}
\alpha = 0.399  \;,  \;\;\;\;\;\;   
\gamma = 0.899  \;,  \;\;\;\;\;\;
\beta  = 0.767  \;\;.
\end{equation}     
The last term in (\ref{Efiicqcad}) with $\delta = 0.4355$ guarantees 
the pure ionic quantum correction in our formalism. 
$D_{\rm 3}(\eta)$ is the Debye integral, which can be approximated 
by (see e.g. Iben \& Tutukov 1984)
\begin{equation}
D_{\rm 3} (\eta)  = \frac{3}{\eta^3} \int ^{\eta}_{0} {\rm d}t
                    \frac{t^3}{{\rm e}^t -1}
  \approx \left ( 1 + 0.43 \eta + \frac{5}{\pi^4} \eta^3\right)^{-1}
 \;\;.
\end{equation}
With (\ref{Efiicqcad}) our set of potentials is given by
\begin{eqnarray}
p_{\rm ii}^{\rm cq} & = & D_{\rm 3}(\alpha\Theta_{\rm i})
         + \frac{1}{2} \; \frac{\gamma\Theta_{\rm i}
                     {\rm {exp}}(-\gamma\Theta_{\rm i})}
               {1 -  {\rm {exp}}(-\gamma\Theta_{\rm i})}
         - \frac{3}{2} +\frac{1}{2}\beta\Theta_{\rm i}
\end{eqnarray}
\begin{eqnarray}
g_{\rm ii}^{\rm cq} & = & p_{\rm ii}^{\rm cq} + f_{\rm ii}^{\rm cq}
\end{eqnarray}
\begin{eqnarray}
u_{\rm ii}^{\rm cq} & = & 2 p_{\rm ii}^{\rm cq}
\end{eqnarray}
\begin{eqnarray}
\frac{1}{k_{\rm T,ii}^{\rm cq}} & = & p_{\rm ii}^{\rm cq}
 -   \frac{3}{2} \; D_{\rm 3} (\alpha \Theta_{\rm i})
 +   \frac{3}{2} \;  \frac{\alpha\Theta_{\rm i}
                {\rm {exp}}(-\alpha\Theta_{\rm i})}
           {1 - {\rm {exp}}(-\alpha\Theta_{\rm i})}
                + \frac{1}{4} \beta\Theta_{\rm i}
      \nonumber \\
 & &  \quad\quad
      + \frac{1}{4} \; \frac{\gamma\Theta_{\rm i}
                {\rm {exp}}(-\gamma\Theta_{\rm i})}
           {1 - {\rm {exp}}(-\gamma\Theta_{\rm i})}
          \left[   1 - \frac{\gamma\Theta_{\rm i}}
           {1 - {\rm {exp}}(-\gamma\Theta_{\rm i})} \right]
      \nonumber \\
 & &  \quad\quad
 +   \frac{3}{2} \;    \frac{\alpha\Theta_{\rm i}
                {\rm {exp}}(-\alpha\Theta_{\rm i})}
           {1 - {\rm {exp}}(-\alpha\Theta_{\rm i})}
\end{eqnarray}
\begin{equation}
\phi_{\rm S,ii}^{\rm cq}  = 3 p_{\rm ii}^{\rm cq} -
                            2 \frac{1}{k_{\rm T,ii}^{\rm cq}}
\end{equation}
\begin{equation}  \label{Ecvcqiiws}
c_{\rm V,ii}^{\rm cq} = 2 \phi_{\rm S,ii}^{\rm cq}
\end{equation}
An explicite expression for (\ref{Ecvcqiiws}) is given by Chabrier (1993).
Figs.~\ref{Ff_qc_ii_T6}-\ref{Fcv_r75_C_nnn_cad}
compare the ionic quantum corrections for the free energy as well as the 
isochoric specific heat calculated by different approaches. 
Note, the end of the solid lines in the 
Figs.~\ref{Ff_qc_ii_T6}-\ref{Fcv_r75_C_nnn_cad} reflects the validity of the 
moment expansion as pointed out by Nagara \etal\ (1987).
Fig.~\ref{Ff_qc_ii_T6} shows ionic quantum corrections for 
$Z = 6$ calculated with different theories.    
Ionic quantum corrections in the vicinity of phase transition 
at $\Gamma_{\rm i} \approx 180$ are shown in Fig.~\ref{Ff_qc_ii_G180}.
Fig.~\ref{Ff_qc_ii_G180} illustrates that the quantum corrections 
calculated by Iyetomi \etal\ (1993) decrease with increasing $\Theta_{\rm i}$. 
This behaviour is  caused by the inclusion of the anharmonic contribution 
of the zero-temperature oscillations 
(see e.g. Nagara \etal\ 1987, Iyetomi \etal\ 1993), 
which is neglected by Chabrier \etal\ (1992).  
Fig.~\ref{Fcv_cqii_r75_C} shows the ionic quantum corrections for the 
isochoric specific heat derived from the corresponding free energies. 
Fig.~\ref{Fcv_r75_C_nnn_cad} illustrates the course of the
electronic and ionic contributions of the isochoric specific heat
vs.\ temperature and shows how different prescriptions for the
quantum corrections affect the results.
The discontinuities in the curves for the ionic contribution in 
Fig.~\ref{Fcv_r75_C_nnn_cad} refer to the fluid-solid phase transition 
at $\Gamma_{\rm i} \approx 180$.    
\subsubsection{Ion-electron interaction}
Next we have to include the screening effect between electrons and ions.
We adopt the new Pad\'e Approximant based on the Uns\"old-Berlin-Montroll
asymptotics of Stolzmann \& Ebeling (1998) and Ebeling \etal\ (1999)
\begin{equation}  \label{Efiec}
f_{\rm ie}^{\rm c} = - \frac{ c_{0} \Gamma_{\rm i}^{3/2} +
                              c_{2} \Gamma_{\rm e}^{5/2} 
    \varepsilon_{\rm ie}(r_{\rm s}, \Gamma_{\rm i}) 
                      + c_{3} c_{4} \Gamma_{\rm e}^{5/2}}
                         {1 + c_{1} \Gamma_{\rm i}^{3/2} 
                            + c_{2} \Gamma_{\rm e}^{5/2}
                            + c_{3} \Gamma_{\rm e}^{5/2}}
\end{equation}
The weak coupling limit ($\Gamma_{\rm i} \ll 1$) is given here by the Debye-
H\"uckel two component plasma (DHTCP) law.
The coefficients $c_{\rm k}$ are defined by
\begin{equation}   \label{Ecoeffc0}
c_{0} = \frac{1}{\sqrt{3}} \frac{ (\zeta_{\rm e} + \langle Z^{2} 
                                                   \rangle)^{3/2} -
              \zeta_{\rm e}^{3/2} -  \langle Z^{2} \rangle ^{3/2}}
           {\langle Z \rangle^{1/2} \langle Z^{5/3} \rangle^{3/2}} \;, \quad
\end{equation}
\begin{figure}
\epsfxsize=0.49\textwidth
\mbox{\epsffile{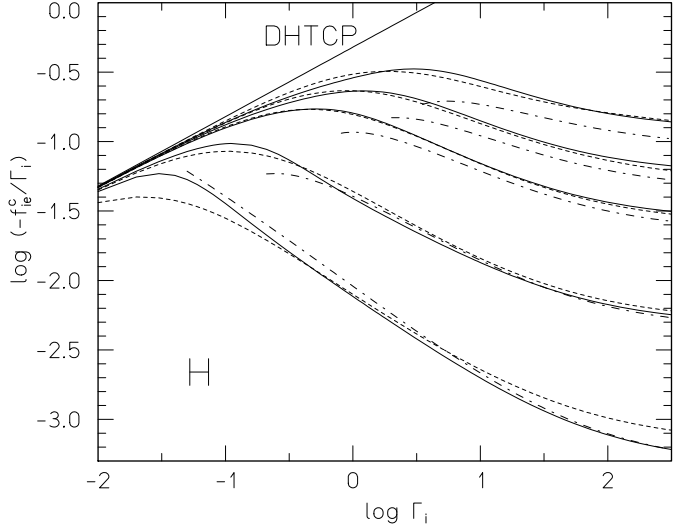}}
\caption[fiec_com_rs_H]
{The free energy contribution of the screening contribution for hydrogen
 and densities referring to $r_{\rm s}=2, 1, 0.5, 0.1, 0.01$ (from top to
 bottom) compared with the Debye-Hueckel limiting law (DHTCP)
 (the first term in the nominator of Eq.~(\ref{Efiec})) and the high
 density/large coupling-asymptotics from Galam \& Hansen (1976)
 (dashed-dotted lines).
 The dashed lines refer to Chabrier \& Potekhin (1998) and the solid lines
 correspond to Eq.~(\ref{Efiec}).}          \label{Ffiec_com_rs_H}
\end{figure}
\begin{figure}
\epsfxsize=0.49\textwidth
\mbox{\epsffile{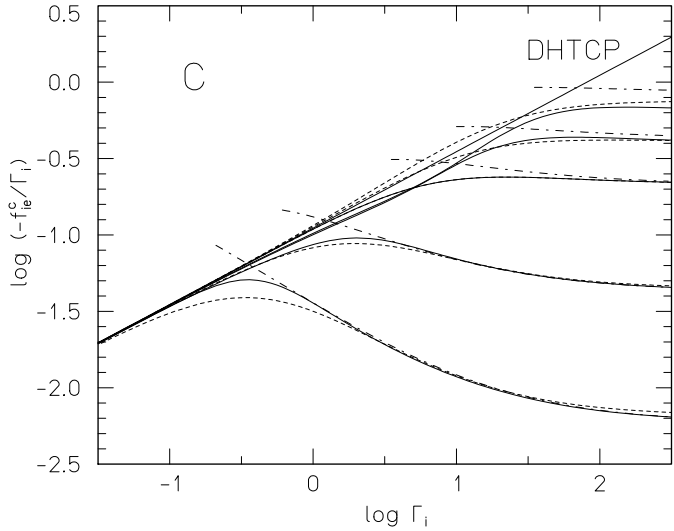}}
\caption[fiec_com_rs_C]
{The free energy contribution of the screening contribution for carbon
 and densities referring to $r_{\rm s}=2, 1, 0.5, 0.1, 0.01$ (from top to
 bottom) compared with the Debye-Hueckel limiting law (DHTCP)
 and the high density/large coupling-asymptotics from
 Galam \& Hansen (1976) (dashed-dotted lines).
 The dashed lines refers to Chabrier \& Potekhin (1998) and the solid lines
 correspond to Eq.~(\ref{Efiec}).}          \label{Ffiec_com_rs_C}
\end{figure}  
\begin{eqnarray}     \label{Ecoeffc1}
c_{1}  & = & \frac{3}{2} \frac{1}{c_{0} \langle Z \rangle
             \langle Z^{5/3} \rangle^{3}}  
  \left[ \frac{\sqrt{\pi\tau}}{8} \zeta_{\rm e} \langle Z^{2} \rangle
   -    \right. \nonumber  \\
   & & \left.
  \sum_{ab}  \zeta_{a} \zeta_{b} Z_{a}^{3}
     Z_{b}^{3}  \left( \frac{Q_{3}( -|\xi_{ab}|)}{|\xi_{ab}|^{3}}
- \frac{1}{6} \ln|\xi_{ab}| - C_{0} \right) \right]
\end{eqnarray}
\begin{equation}
c_{2} = Z r_{\rm s}^{-7/4}
  \; , \quad\quad\quad\quad
c_{3} = 0.01 
  \; , 
\end{equation}
\begin{equation}
c_{4} = \left(2 \cdot 1.786 - 0.9 - 0.9 \Gamma_{\rm e}/\Gamma_{\rm i}
        \right) \Gamma_{\rm i}
\end{equation}
\begin{equation}
\xi_{ab} = - \frac {2 Z_{a}Z_{b}}{\sqrt{\tau(\gamma_{a}+\gamma_{b})}} \; ,
\quad\quad
\gamma_{k} = \frac{m_{\rm e}}{m_{k}} \; ,
\quad\quad
\zeta_{e} = \frac{ n_{e}}{ \sum_{i} n_{i}}  
\end{equation}
In Eq.~(\ref{Ecoeffc1}) is summed up over $a \not= b$ in contrast to summations
made in earlier approaches (Ebeling 1993, paper I).
\begin{figure}
\epsfxsize=0.49\textwidth
\mbox{\epsffile{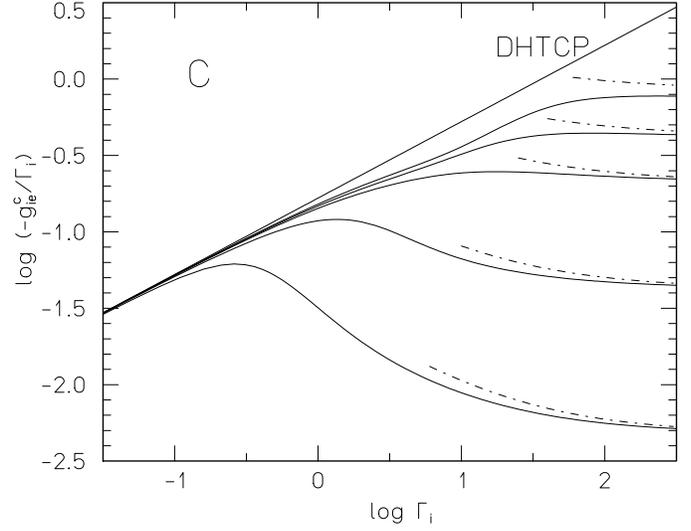}}
\caption[giec_com_rs_C]
{The free gibbs energy contribution of the screening contribution for carbon
 and densities referring to $r_{\rm s}=2, 1, 0.5, 0.1, 0.01$ (from top to
 bottom) compared with the Debye-Hueckel limiting law (DHTCP)
 and the high density/large coupling-asymptotics from
 Galam \& Hansen (1976) (dashed-dotted lines).
 The solid lines correspond to Eq.~(\ref{Egiec}).}
                                      \label{Fgiec_com_rs_C}
\end{figure}  
For $\varepsilon_{\rm  ie}(r_{\rm s}, \Gamma_{\rm i})$
we apply the expression given by Ebeling \& Richert (1985), which consider
the high density/large coupling-asymptotics from Galam \& Hansen (1976)
including high-density corrections according to Kagan \etal\ (1977).
\begin{equation}
 \varepsilon_{\rm ie}(r_{\rm s}, \Gamma_{\rm i}) =
  \left( \frac{12 Z}{\pi^{5/2}} \right)^{2/3} \Gamma_{\rm i} r_{\rm s}     
  \left[ \varepsilon^{0}_{\rm ie}(r_{\rm s}, \Gamma_{\rm i}) 
   +  x^{2} J (\Gamma) \right] 
\end{equation}
\begin{equation}
 \varepsilon^{0}_{\rm ie} =
       \frac{ A + C     \; \Gamma_{\rm i}^{-1}}{1 + 0.089 r_{\rm s}^{2}}
   +   \frac{ B         \; \Gamma_{\rm i}^{-3/4}}
            { 1 + 0.353 \; \Gamma_{\rm i}^{-1/4}}
\end{equation}
with the relativity parameter $x = 0.014 / r_{\rm s}$.
$J (\Gamma_{\rm i})$ as well as $A$, $B$ and $C$ depending on Z are given
by Galam \& Hansen (1976).
Recently, Chabrier \& Potekhin (1998) presented a Pad\'e formula for the 
screening contribution of the free energy, which includes the Thomas-Fermi 
approximation in the limit of high densities and large coupling parameters. 
The results of Chabrier \& Potekhin (1998) with Eq.~(\ref{Efiec}) are compared 
for hydrogen and carbon at five densities in Fig.~\ref{Ffiec_com_rs_H} 
and Fig.~\ref{Ffiec_com_rs_C}. \\ 
In the following we proceed in the same manner as 
in the case of the electron-electron subsystem in Sect.~\ref{eesec}. \\ 
For the ion-electron Gibbs energy we use the Pad\'e expression
\begin{equation}  \label{Egiec}
g_{\rm ie}^{\rm c} = - \frac{ \frac{3}{2} c_{0} \Gamma_{\rm i}^{3/2}    
                                        + c_{2} \Gamma_{\rm e}^{5/2}
  \left(\varepsilon_{\rm ie}+ \frac{1}{3} \left(h_{\rm ie}-d_{\rm ie}\right)
  \right)  
                                  + c_{3} c_{4} \Gamma_{\rm e}^{5/2}}
                                     {1 + c_{1} \Gamma_{\rm i}^{3/2} 
                                        + c_{2} \Gamma_{\rm e}^{5/2}
                                        + c_{3} \Gamma_{\rm e}^{5/2}}
\end{equation}  
with 
$h_{\rm ie}   =  \Gamma_{\rm i} \cdot  \partial \varepsilon_{\rm  ie} / 
 \partial \Gamma_{\rm i}$ and
$d_{\rm ie}  = r_{\rm s} \cdot  \partial \varepsilon_{\rm  ie} /
 \partial r_{\rm s}$.
Fig.~\ref{Fgiec_com_rs_C} illustrates the Gibbbs energy for carbon at five 
densities. 
The polarization pressure can be determined by
$p_{\rm ie}^{\rm c} = g_{\rm ie}^{\rm c} - f_{\rm ie}^{\rm c}$.
\section{Numerical results and discussion}  \label{results}
\begin{figure}
\centering
\epsfxsize=0.49\textwidth
\mbox{\epsffile{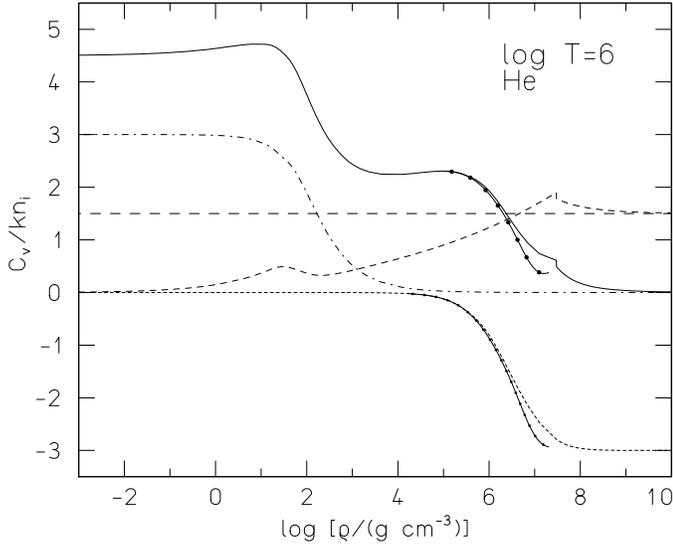}}
\caption[cv_he_6]
{Contributions of the isochoric specific heat in units of $kn_{\rm i}$    
 for helium at $T=10^{6}{\rm K}$ vs.\ density.
 The lines refer to
 the ideal electrons (dashed dotted) and ions (long-dashed),
 the exchange and correlation (dashed),
 the ionic quantum correction from Chabrier \etal\ (1992) (short-dashed), and
 the sum of those (solid). The curves with dots represent ionic quantum 
 corrections after Nagara \etal\ (1987).}
                                                  \label{Fcv_he_6}
\end{figure}
\begin{figure}
\centering
\epsfxsize=0.49\textwidth
\mbox{\epsffile{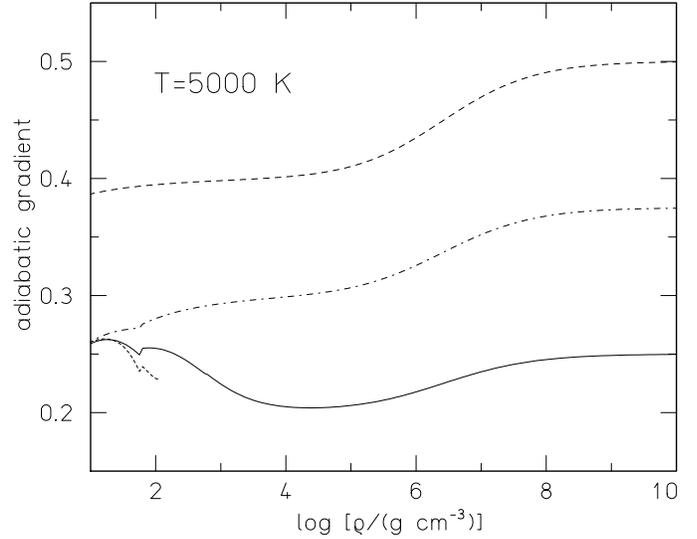}}
\caption[nab_limes2]
{High-density asymptotics of the adiabatic temperature gradient
 $\nabla_{\rm ad}$ vs.\ density
 for hydrogen, demonstrated along the isotherm $T=5 \cdot 10^{3}$ K.
 The dashed line refers to the ideal and exchange terms only.
 The dashed-dotted (without ionic quantum effects), the solid
 (with ionic quantum corrections from Chabrier \etal\ 1992), and the 
 short-dashed (with ionic quantum corrections from Nagara \etal\ 1987) lines 
 take into account correlations. For more explanation: see text.}
                                    \label{Fnab_limes2}
\end{figure}
\begin{figure}
\centering
\epsfxsize=0.49\textwidth
\mbox{\epsffile{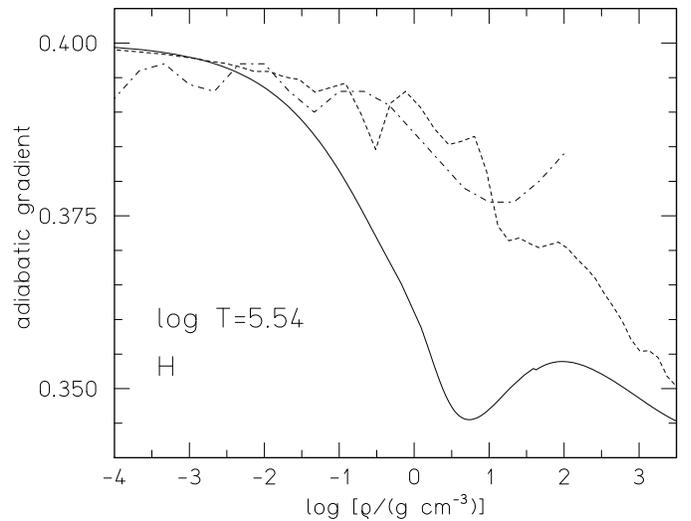}}
\caption[nab_554_H_id]
{Adiabatic temperature gradient $\nabla_{\rm ad}$ (without radiation and
{\it all} terms in Eq.~(\ref{Eicsredu})) for hydrogen (solid line) at   
 $T=10^{5.54}{\rm K}$ vs.\ density.
 The dashed line refers to Saumon \etal\ (1995) and the dashed-dotted line
 to Fontaine \etal\ (1977).}
                                    \label{Fnab_554_H_id}
\end{figure}
\begin{figure}
\centering
\epsfxsize=0.49\textwidth
\mbox{\epsffile{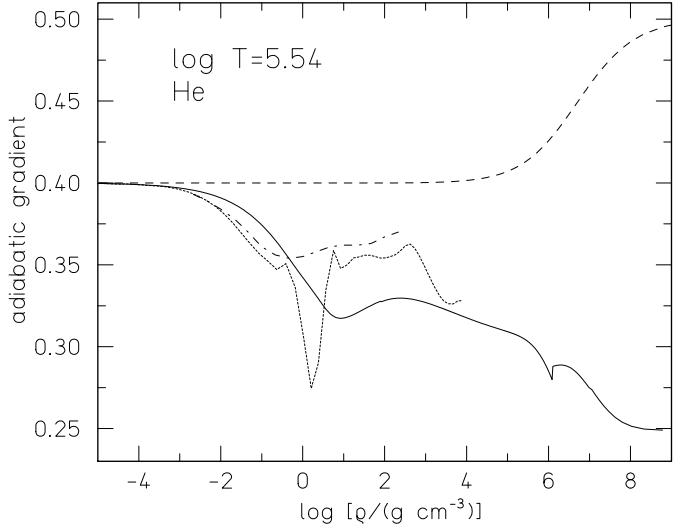}}
\caption[nab_554_He_id]
{Adiabatic temperature gradient $\nabla_{\rm ad}$
 (without radiation and {\it all} terms in Eq.~(\ref{Eicsredu}))
 for helium at $T=10^{5.54}{\rm K}$ vs.\ density.
 For the ionic quantum corrections the results from
 Chabrier \etal\ (1992) and from Nagara \etal\ (1987) (solid line) are used.
 The dotted line refers to Saumon \etal\ (1995) and the dashed-dotted line
 to Fontaine \etal\ (1977).
 The ideality (long-dashed line) is shown for comparison.}
                                   \label{Fnab_554_He_id}
\end{figure}
\begin{figure}
\centering
\epsfxsize=0.49\textwidth
\mbox{\epsffile{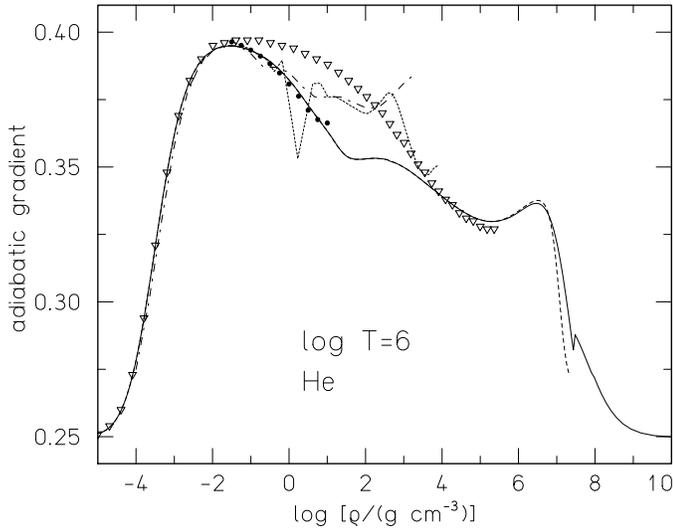}}
\caption[nab_6_He]
{Adiabatic temperature gradient $\nabla_{\rm ad}$
 (with radiation and {\it all} terms in Eq.~(\ref{Eicsredu})) 
 for helium at $T=10^{6}{\rm K}$ vs.\ density. For the ionic quantum 
 corrections the results from Chabrier \etal\ (1992) (solid line)
 and from Nagara \etal\ (1987) (short-dashed line) are used.
 The dotted line refers to Saumon et al.\ (1995), the triangles to
 Straniero (1988), the dashed-dotted line to Fontaine et al.\ (1977),
 and the dots to Rogers et al.\ (1996).
}                                                 \label{Fnab_6_He}
\end{figure}
\begin{figure}
\centering
\epsfxsize=0.49\textwidth
\mbox{\epsffile{0972.f17}}
\caption[nab_6_C]
{Adiabatic temperature gradient $\nabla_{\rm ad}$
 (with radiation and {\it all} terms in Eq.~(\ref{Eicsredu}))
 for carbon (solid line and short-dashed lines) at $T=10^{6}{\rm K}$
 vs.\ density. For the ionic quantum corrections
 the results from Chabrier \etal\ (1992) (solid line)
 and from Nagara \etal\ (1987) (short-dashed line) are used.
 The triangles refers to Straniero (1988), the dashed-dotted line
 to Fontaine et al.\ (1977), and the crosses to Lamb (1974).
}                                                \label{Fnab_6_C}
\end{figure}
\begin{figure}
\centering
\epsfxsize=0.49\textwidth
\mbox{\epsffile{0972.f18}}
\caption[nab_7_C]
{Adiabatic temperature gradient $\nabla_{\rm ad}$
 (with radiation and {\it all} terms in Eq.~(\ref{Eicsredu}))
 for carbon (solid line and short-dashed lines) at $T=10^{7}{\rm K}$
 vs.\ density. For the ionic quantum corrections
 the results from Chabrier \etal\ (1992) (solid line)
 and from Nagara \etal\ (1987) (short-dashed line) are used.
 The triangles refers to Straniero (1988), the dashed-dotted line
 to Fontaine et al.\ (1977), and the crosses to Lamb (1974).
}                                      \label{Fnab_7_C}
\end{figure}
\begin{figure}
\centering
\epsfxsize=0.49\textwidth
\mbox{\epsffile{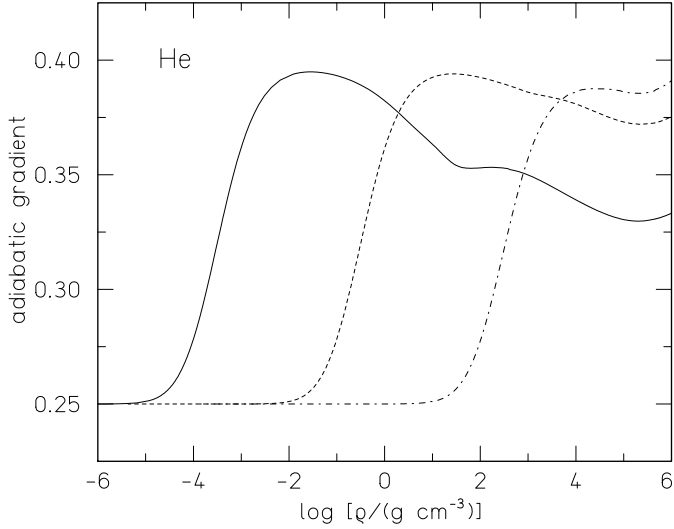}}
\caption[nab_rad_He_th3]
{Adiabatic temperature gradient $\nabla_{\rm ad}$
(with radiation and {\it all} terms in Eq.~(\ref{Eicsredu}))
for helium along the isotherms
$T=10^{6}$ (solid), $10^{7}$ (dashed), and $10^{8}{\rm K}$ (dashed-dotted) 
vs.\ density.}               \label{Fnab_rad_He_th3}
\end{figure}
\begin{figure}
\centering
\epsfxsize=0.49\textwidth
\mbox{\epsffile{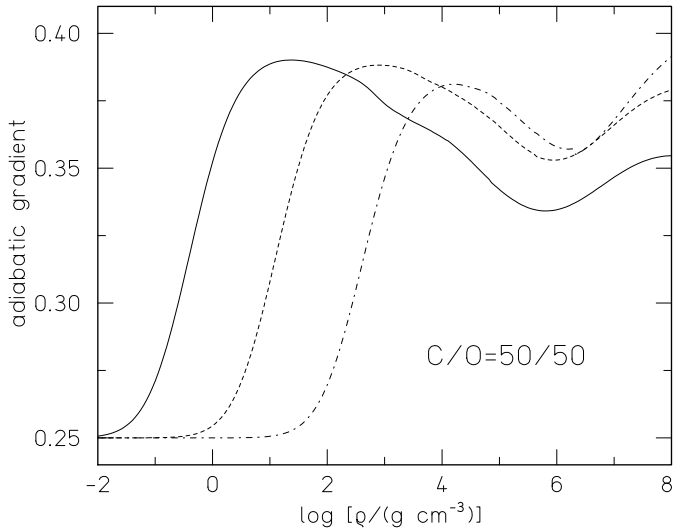}}
\caption[nab_C50O50_Th3]
{Adiabatic temperature gradient $\nabla_{\rm ad}$
(with radiation and {\it all} terms in Eq.~(\ref{Eicsredu})) for a mixture of
C/O=50/50 (mass fractions) along the isotherms
$T=10^{7}$ (solid), $10^{7.5}$ (dashed), and $10^{8}{\rm K}$
(dashed-dotted) vs.\ density.
}                                      \label{Fnab_C50O50_Th3}
\end{figure}
This section will be devoted to illustrate the behaviour of the
adiabatic temperature gradient $\nabla_{\rm ad}$.
The adiabatic temperature gradient is
a highly sensitive quantity, because it depends on first and second-order
derivations of the model Helmholtz free energy.
Therefore it is well suited to test the quality of the explicit expressions 
derived in Sects. 3 and 4.
The adiabatic gradient is a crucial ingredient for many applications.
For instance, $\nabla_{\rm ad}$ determines the Schwarzschild criterion for 
convective instability and an accurate evaluation is required for evolutionary
calculations of very-low mass stars.
We restrict our comparisons of the adiabatic temperature gradient to the 
EOS tables given by Straniero (1988), Saumon \etal\ (1995), and Rogers
\etal\ (1996). Further detailed comparisons can be found in Saumon \etal\ 
(1995) who compared their results for  $\nabla_{\rm ad}$ with those of
Fontaine \etal\ (1977), D\"appen \etal\ (1988), and Magni \& Mazzitelli (1979).

Fig.~\ref{Fcv_he_6} shows the total isochoric specific heat
according to Eq.~(\ref{Eicsredu}) as well as their 
various single contributions (in units of $n_{\rm i} k$), i.e.\
electronic exchange and correlation terms 
(electronic, ionic and screening part).
We note that the discontinuity in the ionic correlation is caused by the
fluid-solid phase transition at $\Gamma=178$ (Stringfellow \etal\ 1990), which
can also be observed for the adiabatic temperature gradient in the
high-density region. The influence of quantum effects on the fluid-solid 
transition of the OCP is discussed in detail by Pollock \& Hansen (1973),
Lamb \& Van Horn (1975), Nagara \etal\ (1987), Chabrier (1993) and 
Iyetomi \etal\ (1993). We chose for our numerical calculations 
$\Gamma=178$, which is based on an simulated OCP without quantum effects 
(Stringfellow \etal\ 1990).   
The specific heat contribution according to the ionic correlation is
monotonously increasing until the phase transition of the OCP model.
The first maximum at $\log \rho \approx 1.5$ in the dashed curve in
Fig.~\ref{Fcv_he_6} is determined essentially by the electronic exchange and
electronic correlation.
The influence of electronic exchange and correlation increases with
decreasing temperatures.
Quantum effects for the specific heat according to the ions after Chabrier 
(1993) are illustrated
by the dotted line in Fig.~\ref{Fcv_he_6}, which can be seen also in
Figs.~\ref{Fcv_cqii_r75_C} and \ref{Fcv_r75_C_nnn_cad}.
This contribution is compensated at very high densities by the   
contributions of the ideal and the correlated ions.
Considering ionic quantum corrections calculated by Nagara \etal\ (1987) 
the isochoric specific heat will be more strongly reduced as shown in
Fig.~\ref{Fcv_he_6}.

We point out that the same behaviour appears for the adiabatic temperature
gradient as indicated in
Figs.~\ref{Fnab_limes2} and \ref{Fnab_554_He_id}-\ref{Fnab_7_C}.
Note, that although the ionic quantum correction calculated by
Chabrier \etal\ (1992) and Chabrier (1993) are valid only for the solid 
phase we considered here also the fluid range in order to illustrate the 
differences to the result of Nagara \etal\ (1987).
\\
Figs.~\ref{Fnab_limes2}-\ref{Fnab_C50O50_Th3}
display isotherms of the adiabatic temperature gradient evaluated
on the basis our analytic expressions given in Sect.~\ref{thermppot} 
including the photonic contribution 
(Figs.~\ref{Fnab_6_He}-\ref{Fnab_C50O50_Th3})
over a broad range of densities.
As already mentioned, the representations towards very high densities
aimed at the purpose to show the asymptotic behaviour of the potentials
given by Eq.~(\ref{Eicsredu}).
The  accurate description of plasmas at very high density region requires  
the consideration of additional effects as electron captures and other 
processes (cf.\ Shapiro \etal\ 1983). 
Fig.~\ref{Fnab_limes2} shows the high density asymptotics of the
adiabatic temperature gradient calculated by different approximations.
Considering ideal and exchange contributions we obtained the well
known value $\nabla_{\rm ad} = 1/2$, according to relativistic considerations. 
The limiting value $\nabla_{\rm ad} = 3/8$ is determined by inclusion 
of the classical ionic correlation and $\nabla_{\rm ad} = 1/4$ 
is obtained with ionic quantum corrections given by Chabrier \etal\ (1992). 
In Fig.~\ref{Fnab_554_H_id}-\ref{Fnab_7_C} we compare our
formalism with data from other authors
(Lamb 1974, Fontaine \etal\ 1977, Straniero 1988,
 Saumon \etal\ 1995, Rogers \etal\ 1996) based on
analytical as well as numerical methods and valid for various density and
temperature ranges.
The radiation contribution is dominant at high temperatures and low densities.
The well-known limiting case of $\nabla_{\rm ad}=0.25$ in the low-density limit
depends on the temperature in a strong manner.

The largest deviations for the adiabatic temperature gradient are observed
in the intermediate density region, that means for plasma parameter where e.g.
correlation effects must be described by accurate expressions.
It should be emphasized that the Pad\'e technique is well suited to
meet the challenging requirements of this - sometimes poorly known - density
regime.

Fig.~\ref{Fnab_rad_He_th3} shows the adiabatic temperature gradient for helium.
In Fig.~\ref{Fnab_C50O50_Th3} we draw $\nabla_{\rm ad}$ for a mixture of
carbon and oxygen at high temperatures as appropriate for white dwarf
interiors.

The temperatures and densities are chosen here according to the conditions
for a fully ionized plasma. Nevertheless, our EOS-formalism can be applied
for the conditions of a partially ionized plasma (lower temperatures)
provided additional terms in Eq.~(\ref{Eicsredu}) are considered and are
solved in connection with the formalism of the dissoziation- and ionization
equilibrium (Beule \etal\ 1999).
\section{Summary}     \label{summ}
Our aim was to provide data for the thermodynamical potentials 
summarized in Eq.~(\ref{Eicsredu}) which are applicable 
to, e.g., stellar interiors (fully ionized regions). 
For the Helmholtz free energy, the Gibbs free energy, and the pressure 
(equation of state), expressions for each term in Eq.~(\ref{Eicsredu}), 
i.e. ideality, exchange and correlation, have already been presented in 
paper I.
In this paper we have improved our Pad\'e Approximants for the correlation 
contributions and continued to present formulae for the explicit 
calculation of the internal energy, reciprocal compressibility (bulk modulus), 
coefficient of strain, and the specific heat.
Further results using our calculational concept for the adiabatic temperature
gradient for fully ionized plasmas are presented by 
Stolzmann \& Bl\"ocker (1998).  The extension of our formalism to the
partially ionized region is in progress. 
\acknowledgements
One of us (W.S.) acknowledges funding by the 
{\it Deutsche Forschungsgemeinschaft}
(grant Scho 394/18).
\end{document}